\documentclass[%
 preprint,
 amsmath,amssymb,
 aps,
]{revtex4-2}
\usepackage[english]{babel} % English and German language 
\usepackage{graphicx}% Include figure files
\graphicspath{{figures/}}
\usepackage{dcolumn}% Align table columns on decimal point
\usepackage{bm}% bold math
\usepackage{subfigure}

\usepackage{amsmath}
\usepackage{tikz}
\usepackage{tikz-cd}
\usetikzlibrary{positioning,arrows} % Adding libraries for arrows
\usetikzlibrary{decorations.pathreplacing} % Adding libraries for decorations and paths
\usepackage{tikzsymbols} % For amazing symbols ;)

\usepackage{tikz}
\usetikzlibrary{shapes,arrows}
\usepgflibrary{arrows}
\usetikzlibrary{decorations.pathreplacing}
\usetikzlibrary{spy}
\usetikzlibrary{shapes,arrows}
\usetikzlibrary{calc}
\usetikzlibrary{arrows, arrows.meta}
\usepgflibrary{arrows}

\usetikzlibrary{shapes.geometric, arrows}
\tikzstyle{process} = [rectangle, minimum width=1.5cm, minimum height=1cm, text centered, text width=1.5cm, draw=black, thick, fill=black!30]
\tikzstyle{whiteprocess} = [rectangle, minimum width=1.5cm, minimum height=0.5cm, text centered, text width=2.1cm, draw=black, thick, fill=black!0]
\tikzstyle{decision} = [diamond, minimum width=4cm, minimum height=1cm, text centered, text width=3cm, draw=black, fill=black!20]
\tikzstyle{arrow} = [thick,->,>=stealth]

\usepackage{relsize}

\tikzset{fontscale/.style = {font=\relsize{#1}}
    }
    
\usepackage[export]{adjustbox}

\usepackage{scalerel}
\usepackage{stackengine,wasysym}
\newcommand\reallywidetilde[1]{\ThisStyle{%
  \setbox0=\hbox{$\SavedStyle#1$}%
  \stackengine{-.1\LMpt}{$\SavedStyle#1$}{%
    \stretchto{\scaleto{\SavedStyle\mkern.2mu\AC}{.5150\wd0}}{.6\ht0}%
  }{O}{c}{F}{T}{S}%
}}
\begin{document}

% \preprint{APS/123-QED}

\title[A large eddy simulation model for two-way coupled particle-laden turbulent flows]{A large eddy simulation model for two-way coupled particle-laden turbulent flows}% Force line breaks with \\

\author{M. Hausmann}
\author{F. Evrard}
\altaffiliation[Also at]{ Sibley School of Mechanical and Aerospace Engineering, Cornell University, Ithaca, NY 14853, USA}
\author{B. van Wachem}
\email{berend.vanwachem@ovgu.de}
\affiliation{ 
Chair of Mechanical Process Engineering, Otto-von-Guericke-Universit{\"a}t Magdeburg, \\
  Universit{\"a}tsplatz 2, 39106 Magdeburg, Germany%\\This line break forced with \textbackslash\textbackslash
}

\date{\today}

\begin{abstract}
In this paper we propose a new modeling framework for large eddy simulations (LES) of particle-laden turbulent flows that captures the interaction between the particle and fluid phase on both the resolved and subgrid-scales. Unlike the vast majority of existing subgrid-scale models, the proposed framework does not only account for the influence of the sugrid-scale velocity on the particle acceleration but also considers the effect of the particles on the turbulent fluid flow. This includes the turbulence modulation of the subgrid-scales by the particles, which is taken into account by the modeled subgrid-scale stress tensor, and the effect of the unresolved particle motion on the resolved flow scales. Our new modeling framework combines a recently proposed model for enriching the resolved fluid velocity with a subgrid-scale component, with the solution of a transport equation for the subgrid-scale kinetic energy. We observe very good agreement of the particle pair separation and particle clustering compared to the corresponding direct numerical simulation (DNS). Furthermore, we show that the change of subgrid-scale kinetic energy induced by the particles can be captured by the proposed modeling framework. 
\end{abstract}

\maketitle

\section{Introduction}
\label{sec:Introduction}
To capture particle-turbulence interactions within the whole turbulence spectrum down to the Kolmogorov length and time scale, a direct numerical simulation (DNS) has to be preformed. For academic cases, DNS of particle-laden turbulent flow is commonly used to gain insights in the underlying physical phenomena. As the flow configurations become more relevant for practical applications, however, resolving such a wide range of flow length and time scales becomes prohibitively expensive. An established surrogate in single phase flows is large eddy simulation (LES), which resolves only the large flow scales and models the mainly dissipative effect of the small scales. While this works well for the majority of single phase flow applications, severe problems can arise if particle-laden flows are considered, especially if the particles significantly affect the flow (two-way coupling). \\
The majority of the literature on particle-laden flows focuses on the one-way coupling, where the modification of the particle statistics by the flow is considered but not the effect of the particles on the flow. The effect of neglecting the subgrid-scale velocity contributions on the transport of the particles has been investigated and quantified in a variety of studies \cite{Armenio1999,Ray2011,Fede2006a,Rosa2017}. It has been observed, that even though the kinetic energy of the subgrid-scale velocity is small, the consequence of neglecting the subgrid-scale fluid velocity on the motion of the particles can strongly affect the preferential concentration and other statistics of the particles. \\
There are several classes of models that attempt to produce realistic particle statistics in the scope of LES. Lagrangian models typically rely on the solution of a stochastic differential equation for every individual particle (see, e.g.,\cite{Fede2006,Bini2007,Berrouk2007,Shotorban2006,Pozorski2009,Knorps2021}). These models are typically simple to implement, computationally efficient and can also be applied in complex domains. However, they usually contain empirical parameters and their Lagrangian nature prevents them from predicting accurate particle pair statistics. Other models rely on successive deconvolution of the LES velocity and use the resulting fluid velocity field to transport the particles \cite{Kuerten2006,Shotorban2005}. Park et al. \cite{Park2017} extended the deconvolution of the filtered velocity, by dynamically adjusting an elliptic differential filter such that the model is either kinetic energy or dissipation consistent with the subgrid-scale model. The main drawback of approximate deconvolution methods is that they do not introduce velocities with higher wave numbers than the LES, but only modify the LES velocity. That is why these models are not able to reproduce the particle statistics of a DNS to the full extent. \\
The most promising predictions of particle statistics using a LES framework are obtained with models that reconstruct the subgrid-scale velocity. Bassenne et al. \cite{Bassenne2019} proposed such a model that alternatingly applies the dynamic approximate deconvolution of Park et al. and a spectral extrapolation based on the work of Domaradzki and Loh \cite{Domaradzki1999}. The proposed model improves the prediction of particle clustering for a wide range of Stokes numbers. However, the model requires a projection operation that has to be carried out with a resolution comparable to the DNS in order to obtain a divergence free subgrid-scale velocity, which introduces prohibitively high computational costs. The kinematic simulation is a much cheaper approach and relies on the reconstruction of the subgrid-scale velocity using a truncated Fourier-series \cite{Ray2014,Zhou2020b,Murray2016}. The Fourier-coefficients required for the kinematic simulation are chosen such that the resulting velocity field is divergence free and matches a given kinetic energy spectrum. Even though kinematic simulations yield improved predictions of particle clustering for Stokes numbers $St>1{.}5$, it only applies to spatially homogeneous problems. Considering the available models, there still does not exist a model that yields satisfying improvements in predicting particle clustering and the Lagrangian particle statistics, while maintaining important features for practical applicability, such as a reasonable computational cost and the absence of empirical model parameters of critical influence. \\
Extensive research has been carried out to better understand the modulation of turbulence by particles. Studies of forced and decaying homogeneous isotropic turbulence (HIT) have shown that the presence of particles can modify the total dissipation in two ways \cite{Mallouppas2017,Ferrante2003,Letournel2020,Boivin1998,Nabavi2022}: (\romannumeral 1) The particles can remove from or add kinetic energy to the turbulent flow. The sign of the particle kinetic energy transfer and the scales at which the kinetic energy transfer occurs have been shown to depend on at least three parameters: the Stokes number, the particle number density, and the mass fraction \cite{Letournel2020}. (\romannumeral 2) The fluid dissipation is influenced by the presence of particles \cite{Mallouppas2017}. Similarly, depending on the characteristic turbulence and particle parameters and on the considered length scales, the fluid dissipation can either be enhanced or diminished. \\
A LES only resolves part of the kinetic energy spectrum and can thus only account directly for a modified total dissipation at the resolved scales. While the particle dissipation at the subgrid-scales is fully disregarded in a classical LES, the subgrid-scale fluid dissipation is assumed to be equal to the fluid dissipation of a single phase flow. Classical LES use one of the many subgrid-scale models designed for single phase flows (see, e.g., \cite{Sagaut2005}) and a fluid-particle coupling force obtained without information of the subgrid-scale fluid velocity at the particle positions. The application of several single phase subgrid-scale models to particle-laden flows has been investigated by Boivin et al. \cite{Boivin2000}, displaying very different results between the models. Boivin et al. also argue that at high particle mass fractions, the modeling error in predicting the fluid dissipation becomes less important, since the particle dissipation is  then dominant. In fact, we shown in the present work that the neglected portion of the particle dissipation and fluid dissipation partially compensate each other. Rohilla et al. \cite{Rohilla2022} show that the Smagorinsky or dynamic Smagorinsky model applied to particle-laden flows is unable to predict the critical particle volume loading at which the turbulence in a channel flow collapses (i.e., the flow becomes laminar). They state that one of the main reasons for this issue is the error made in modeling subgrid-scale dissipation. \\
Due to the complexity of two-way coupled turbulent particle-laden flows, models that account for all the coupling effects between the particles and all fluid length and time scales are very rare. An attempt has been made by Yuu et al. \cite{Yuu2001}, who derived an algebraic model for the subgrid-scale kinetic energy that serves as input for a turbulent viscosity. In the studies of Pannala and Menon \cite{Pannala1999} and Sankaran and Menon \cite{Sankaran2002}, a source term accounting for the presence of particles is added to an evolution equation for the subgrid-scale kinetic energy equation, in a manner somewhat similar to the method presented in this paper. However, their particle source term is not closed, because it contains the subgrid-scale velocity, which requires additional modeling. \\
In this paper we present a framework that accounts for particle turbulence interactions that are typically neglected in a LES. The framework contains two coupled models: (\romannumeral 1) a subgrid-scale model based on the localized dynamic kinetic energy model (LDKM) of Menon and coworkers \cite{Kim1999, Menon1996} with an additional source term accounting for the influence of the particles on the subgrid-scale kinetic energy and (\romannumeral 2), a model for the subgrid-velocity that is used to close the particle equations of motion and the particle source terms in the momentum and subgrid-scale kinetic energy equations \cite{Hausmann2022a}. In section \ref{sec:Theory}, the general numerical framework for treating the particle-laden flows in this paper is introduced, including the transport equation for the subgrid-scale kinetic energy upon which the subgrid-scale model is built.
Section \ref{sec:Modeling} gives an overview of the closures that are required in a particle-laden LES and provides a derivation of the proposed modeling framework. Subsequently, simulation setups for one-way and two-way coupled HIT are introduced in section \ref{sec:Simulations}, and results of the comparison between DNS, LES and modeled LES are presented in section \ref{sec:Results}. Finally, section \ref{sec:Conclusions} concludes the present paper.

\section{General numerical framework}
\label{sec:Theory}
In this work we consider an incomressible fluid of density $\rho_{\mathrm{f}}$ and kinematic viscosity $\nu_{\mathrm{f}}$ in the absence of a gravitational field. The fluid is laden with particles of index $p$ having a density $\rho_p$ and volume $V_p$. By volume filtering the Navier-Stokes equations (NSE), the effect of the particles on the fluid can be modeled without needing to solve for the detail of the flow around each individual particle. The following equations are commonly used to approximate the volume filtered velocity $\boldsymbol{u}$ and pressure $p$ for small particle volume fractions (see, e.g., \cite{Maxey2017}):
\begin{align}
\label{eq:NSE1}
    \nabla \cdot \boldsymbol{u} &= 0, \\
\label{eq:NSE2}
    \dfrac{\partial \boldsymbol{u}}{\partial t} + \nabla \cdot ( \boldsymbol{u}\otimes\boldsymbol{u}) &= -\dfrac{1}{\rho_{\mathrm{f}}}\nabla p + \nabla\cdot\boldsymbol{\sigma} -\dfrac{1}{\rho_{\mathrm{f}}}\sum_p g(|\boldsymbol{x}-\boldsymbol{x}_p|) \boldsymbol{F}_{p},
\end{align}
where $\boldsymbol{\sigma} = \nu_{\mathrm{f}}(\nabla \boldsymbol{u} + (\nabla\boldsymbol{u})^{\mathrm{T}})$ is the viscous stress tensor, $\boldsymbol{F}_p$ is the force contribution of the particle with the index $p$, and $g$ is the kernel of the volume filtering operation (see, e.g., Anderson and Jackson \cite{Anderson1967}) with a filter size $\delta$. Note that the filtering is only applied over volumes that are occupied by the fluid. Strictly speaking the solution of equations \eqref{eq:NSE1} and \eqref{eq:NSE2} is the approximation of volume filtered quantities, which is not equivalent to the actual fluid velocity and pressure. In a simulation, the smallest resolvable flow structures are related to the smallest affordable cell size of the numerical grid. Similar to the LES approach, the volume-filtered approach solves for the large scales that can be resolved by the grid and models the effect of the small scales. \\
Particles are considered as Lagrangian rigid point-particles. The particle position $\boldsymbol{x}_p$ is governed by the equation
\begin{align}
\label{eq:particleeom1}
    \dfrac{\mathrm{d} \boldsymbol{x}_p}{\mathrm{d} t} = \boldsymbol{v}_p,
\end{align}
and the particle velocity by Newton's second law
\begin{align}
\label{eq:particleeom2}
    \dfrac{\mathrm{d} \boldsymbol{v}_p}{\mathrm{d} t} = \dfrac{1}{\rho_p V_p} \boldsymbol{F}_p,
\end{align}
where $\boldsymbol{F}_p$ the sum of all forces acting on the particle. There is a variety of mechanisms that lead to different forces acting on the particle. A summary of the possible force contributions and the regimes whereby their consideration is important has been provided by Kuerten \cite{Kuerten2016}.\\
We only consider cases, in which the particles are significantly smaller than the Kolmogorov length scale (i.e., the smallest turbulent structures). The turbulence of the scales down to the Kolmogorov length scale are resolved with the DNS. Note that the DNS is based on the assumption of point-particles. In cases, where not even the smallest flow structures can be resolved by the numerical grid, a LES can be performed. The governing equations for the LES are obtained by filtering equations \eqref{eq:NSE1} and \eqref{eq:NSE2} once more with a filter $G$ of width $\Delta$, with $\Delta\gg\delta$:
\begin{align}
\label{eq:LESNSE1}
    \nabla \cdot \tilde{\boldsymbol{u}} &= 0, \\
\label{eq:LESNSE2}
    \dfrac{\partial \tilde{\boldsymbol{u}}}{\partial t} + \nabla \cdot ( \widetilde{\boldsymbol{u}\otimes\boldsymbol{u}}) &= -\dfrac{1}{\rho_{\mathrm{f}}}\nabla \tilde{p} + \nabla\cdot \tilde{\boldsymbol{\sigma}} -\dfrac{1}{\rho_{\mathrm{f}}}\sum_p \reallywidetilde{g(|\boldsymbol{x}-\boldsymbol{x}_p|) \boldsymbol{F}_{p}}
\end{align}
Note that the filter $G$ is applied to already continuous quantities (due to the previous filtering with $g$) and $\Tilde{.}$ represents the filtering operator. No further assumptions are introduced with the second filter level. \\
Due to the fact that the particles are much smaller than the Kolmogorov length scale (and therefore of the grid spacing), the numerical discretization of the source terms is realized with the particle-source-in-cell (PSIC) method of Crowe et al. \cite{Crowe1977}:
\begin{align}
\label{eq:PSIC}
    \sum_p g(|\boldsymbol{x}-\boldsymbol{x}_p|) \boldsymbol{F}_{p} \approx \dfrac{1}{V_{\mathrm{cell}}(\boldsymbol{x})} \sum_{p \in \Omega_{\mathrm{cell}}(\boldsymbol{x})} \boldsymbol{F}_{p},
\end{align}
and 
\begin{align}
\label{eq:PSICLES}
    \sum_p \reallywidetilde{g(|\boldsymbol{x}-\boldsymbol{x}_p|) \boldsymbol{F}_{p}} \approx \dfrac{1}{\tilde{V}_{\mathrm{cell}}(\boldsymbol{x})} \sum_{p \in \tilde{\Omega}_{\mathrm{cell}}(\boldsymbol{x})} \boldsymbol{F}_{p},
\end{align}
where $\Omega_{\mathrm{cell}}$ and $\tilde{\Omega}_{\mathrm{cell}}$ indicate computational cells of the DNS and the LES, respectively, and $V_{\mathrm{cell}}<\tilde{V}_{\mathrm{cell}}$ their volumes. \\
For the modeling of the flow scales that are filtered out by the kernel $G$, a transport equation for the subgrid-scale kinetic energy $K_{\mathrm{sgs}} = 1/2 (\widetilde{\boldsymbol{u} \cdot \boldsymbol{u}} - \tilde{\boldsymbol{u}} \cdot \tilde{\boldsymbol{u}})$ is derived. This is done in two steps. Firstly, equation \eqref{eq:NSE2} is dotted with the velocity $\boldsymbol{u}$ and subsequently filtered with $G$, which yields 
\begin{align}
\label{eq:filteredKE}
   \dfrac{1}{2} \dfrac{\partial\widetilde{\boldsymbol{u} \cdot \boldsymbol{u}}}{\partial t} + \dfrac{1}{2}\nabla \cdot \left( \reallywidetilde{ \boldsymbol{u}\otimes\boldsymbol{u} \cdot \boldsymbol{u}} \right) = -\dfrac{1}{\rho_{\mathrm{f}}} \nabla \cdot (\reallywidetilde{p\boldsymbol{u}}) + \nabla (\reallywidetilde{\boldsymbol{\sigma} \cdot \boldsymbol{u} }) - \reallywidetilde{\nabla \boldsymbol{u} : \boldsymbol{\sigma}}\nonumber \\
   -\dfrac{1}{\rho_{\mathrm{f}} \tilde{V}_{\mathrm{cell}}}\sum_{p \in \tilde{\Omega}_{\mathrm{cell}}} \boldsymbol{F}_{p}(\boldsymbol{u}(\boldsymbol{x}_p))\cdot \boldsymbol{u}(\boldsymbol{x}_p).
\end{align}
The last term on the right-hand side is realized by multiplying the drag acting on a particle with the fluid velocity at the particle position and taking the sum over all particles within a LES grid cell. The sum over all particles within a LES grid cell replaces the filtering operation $\reallywidetilde{\boldsymbol{F}_{p}\cdot \boldsymbol{u}}$. In fact, this resembles the approach of Schumann \cite{Schumann1975}, who defines a set of LES equations based on averaging over the volume of the computational cell, which is arguably closer to the numerical realization of a LES than a spatially continuous filtering operation. Secondly, equation \eqref{eq:LESNSE2} is dotted with the filtered velocity $\tilde{\boldsymbol{u}}$, which leads to 
\begin{align}
\label{eq:KEfiltered}
    \dfrac{1}{2}\dfrac{\partial \tilde{\boldsymbol{u}} \cdot \tilde{\boldsymbol{u}}}{\partial t} + \dfrac{1}{2}\nabla \cdot ( \widetilde{\boldsymbol{u}\otimes\boldsymbol{u}}\cdot\tilde{\boldsymbol{u}}) =
    -\dfrac{1}{\rho_{\mathrm{f}}}\nabla \cdot ( \widetilde{p}\tilde{\boldsymbol{u}}) + \nabla\cdot( \widetilde{\boldsymbol{\sigma}}\cdot\tilde{\boldsymbol{u}}) - \nabla \tilde{\boldsymbol{u}}:\widetilde{\boldsymbol{\sigma}}\nonumber \\
    -\dfrac{1}{\rho_{\mathrm{f}} \tilde{V}_{\mathrm{cell}}}\tilde{\boldsymbol{u}}\cdot\sum_{p \in \tilde{\Omega}_{\mathrm{cell}}} \boldsymbol{F}_{p}(\boldsymbol{u}).
\end{align}
The numerical realization of the last term on the right-hand side includes a sum of the forces acting on all particles within a LES grid cell and subsequent multiplication with the LES velocity of the present grid cell. \\
Subtracting equation \eqref{eq:KEfiltered} from equation \eqref{eq:filteredKE} yields an equation for $K_{\mathrm{sgs}}$
\begin{align}
\label{eq:Ksgs}
    \dfrac{\partial K_{\mathrm{sgs}}}{\partial t} + \dfrac{1}{2}\dfrac{\partial}{\partial x_j}(\reallywidetilde{u_i u_j u_i} - \widetilde{u_i u_j}\tilde{u}_i) = -\dfrac{1}{\rho_{\mathrm{f}}}\dfrac{\partial}{\partial x_i}(\widetilde{p u_i} - \tilde{p}\tilde{u}_i) + \nu_{\mathrm{f}} \dfrac{\partial^2 K_{\mathrm{sgs}}}{\partial x_j \partial x_j} \nonumber \\
    -\nu_{\mathrm{f}} \left( \reallywidetilde{\dfrac{\partial u_i}{\partial x_j}\dfrac{\partial u_i}{\partial x_j}} - \dfrac{\partial \tilde{u}_i}{\partial x_j}\dfrac{\partial \tilde{u}_i}{\partial x_j}\right) - \dfrac{1}{\rho_{\mathrm{f}} \tilde{V}_{\mathrm{cell}}}\left(\sum_{p \in \tilde{\Omega}_{\mathrm{cell}}} F_{i,p}(\boldsymbol{u}(\boldsymbol{x}_p)) u_i(\boldsymbol{x}_p) - \tilde{u}_i\sum_{p \in \tilde{\Omega}_{\mathrm{cell}}} F_{i,p}(\boldsymbol{u}(\boldsymbol{x}_p))\right).
\end{align}
The last term on the right-hand side is the source term due to the subgrid-scale kinetic energy transfer by the particles. This equation is the foundation for the modeling of the turbulence modulation by particles of scales that are filtered out by $G$.

\section{Modeling in the LES framework}
\label{sec:Modeling}
The second filtering operation with $G$ leads to equations governing the fluid behavior that can not be solved explicitly without knowing the fluid velocity $\boldsymbol{u}$. We model these equations in the framework of a LES. In addition to the modeling of the subgrid-scale stress tensor, a particle-laden flow requires further closures, which are first explained and then modeled.

\subsection{Required closures}
The fluid equations with the assumption of a dilute particle-laden flow with sufficiently small particles and subsequent filtering with the kernel $G$ can be written in the typical form of a LES \cite{Mallouppas2013b} as
\begin{align}
\label{eq:LES1}
    \dfrac{\partial \tilde{u}_i}{\partial x_i} &= 0, \\
\label{eq:LES2}
    \dfrac{\partial \tilde{u}_i}{\partial t} + \tilde{u}_j \dfrac{\partial \tilde{u}_i}{\partial x_j} &= -\dfrac{1}{\rho_f} \dfrac{\partial \tilde{p}}{\partial x_i} + \nu_{\mathrm{f}} \dfrac{\partial^2 \tilde{u}_i}{\partial x_j \partial x_j} - \dfrac{\partial \tau_{ij}}{\partial x_j} - \dfrac{1}{\rho_{\mathrm{f}} \tilde{V}_{\mathrm{cell}}} \sum_{p \in \tilde{\Omega}_{\mathrm{cell}}}F_{\mathrm{D},i,p}(\boldsymbol{u}(\boldsymbol{x}_p)),
\end{align}
where we assumed that only the drag force $\boldsymbol{F}_{\mathrm{D}}$ acts on the particles. The subgrid-scale stress tensor $\tau_{ij}$ is defined as
\begin{align}
    \tau_{ij} = \widetilde{u_i u_j} - \tilde{u}_i \tilde{u}_j.
\end{align}
With the particle force reducing to the drag force, the particle velocity $\boldsymbol{v}_p$ is governed by
\begin{align}
    \dfrac{\mathrm{d} \boldsymbol{v}_p}{\mathrm{d} t} = \dfrac{1}{\rho_p V_p}\boldsymbol{F}_{\mathrm{D},p}(\boldsymbol{u}(\boldsymbol{x}_p)).
\end{align}
In order to solve for the filtered fluid velocity $\tilde{\boldsymbol{u}}$ and the particle velocity $\boldsymbol{v}_p$, two further modeling steps are required: (\romannumeral 1) The subgrid-scale stress tensor $\tau_{ij}$ has to be closed to model the effect of the unresolved subgrid-scales on the resolved (filtered) quantities. It should be noted that the presence of the particles modifies the subgrid-scale velocity. As a consequence, models for the subgrid-scale stress tensor that are based on assumptions of the single-phase flow subgrid-scale velocity are, strictly speaking, not valid. (\romannumeral 2) In order to compute the drag force acting on the particle and, vice versa, the force that is coupled back to the fluid with opposite sign, the unfiltered fluid velocity at the particle position is required, which is an unknown quantity in a LES. It is well understood, that the particles behave differently when the drag force is obtained from the filtered fluid velocity at the particle position $\boldsymbol{F}_{\mathrm{D},p}(\tilde{\boldsymbol{u}}(\boldsymbol{x}_p))$ \cite{Ray2011,Fede2006a}. In the present paper, a modeling framework for both closures is provided.

\subsection{Modeling the subgrid-scale stress tensor}
\label{ssec:modelingtau}
The subgrid-scale stress tensor accounts for the effect of the subgrid-scale velocities on the velocity that is resolved in a LES. In a LES this subgrid-scale stress tensor is modeled. Typically, the focus lies exclusively on modeling the energetic effects of the subgrid-scales on the resolved scales, even though the mechanisms of turbulent energy transfer (vorticity stretching and strain self-amplification) possess characteristic directional dependencies \cite{Johnson2021}. In single phase turbulent flows, the construction of the subgrid-scale model can be based on the assumption that the energy transferred towards smaller scales is either dissipated by the viscosity or scattered back towards larger scales. In a particle-laden turbulent flow, however, additional energy sources and sinks occur due to the interactions with the particles that classical subgrid-scale models (designed for single phase flows) do not account for. \\
In order to take the interactions of the fluid with the particles into account, we modify the localized dynamic kinetic energy model (LDKM) of Menon and coworkers \cite{Kim1999, Menon1996}. The modeling of the subgrid-scale stress tensor is based on an eddy viscosity $\nu_{\mathrm{k}}$
\begin{align}
    \tau_{ij} = -2 \nu_{\mathrm{k}} \tilde{S}_{ij} + \dfrac{2}{3}K_{\mathrm{sgs}} \delta_{ij},
\end{align}
where $\tilde{S}_{ij}$ is the filtered strain-rate tensor and $\delta_{ij}$ is the Kronecker tensor. The eddy viscosity is computed from the subgrid-scale kinetic energy
\begin{align}
    \nu_{\mathrm{k}}=C_{\mathrm{k}} \sqrt{K_{\mathrm{sgs}}}\Delta,
\end{align}
where $C_{\mathrm{k}}$ is a constant. The evolution of the subgrid-scale kinetic energy $K_{\mathrm{sgs}}$ is governed by the transport equation \eqref{eq:Ksgs}. However, several terms of the transport equation for $K_{\mathrm{sgs}}$ require the unfiltered fluid velocity. Lilly \cite{Lilly1966} introduce a model for the transport equation \eqref{eq:Ksgs} (without particle source term), such that it can be solved by knowing filtered quantities only, which is the basis of the LDKM
\begin{align}
\label{eq:modeledKsgs}
    \dfrac{\partial K_{\mathrm{sgs}}}{\partial t} + \tilde{u}_i \dfrac{\partial K_{\mathrm{sgs}}}{\partial x_i} = -\tau_{ij}\dfrac{\partial \tilde{u}_i}{\partial x_j} - C_{\epsilon}\dfrac{K_{\mathrm{sgs}} ^{3/2}}{\Delta} + \dfrac{\partial }{\partial x_i}\left( \nu_{\mathrm{k}} \dfrac{\partial K_{\mathrm{sgs}}}{\partial x_i}\right)+ \varPhi_{\mathrm{P}},
\end{align}
where $C_{\epsilon}$ is a constant. The particle source term $\varPhi_{\mathrm{P}}$ is not part of the original LDKM but introduced in the present model based on the derivations in section \ref{sec:Theory}:
\begin{align}
\label{eq:ParticleSourceTerm}
     \varPhi_{\mathrm{P}} = - \dfrac{1}{\rho_{\mathrm{f} }\tilde{V}_{\mathrm{cell}}}\left(\sum_{p \in \tilde{\Omega}_{\mathrm{cell}}} F_{\mathrm{D},i,p}(\boldsymbol{u}(\boldsymbol{x}_p)) u_i(\boldsymbol{x}_p) - \tilde{u}_i\sum_{p \in \tilde{\Omega}_{\mathrm{cell}}} F_{\mathrm{D},i,p}(\boldsymbol{u}(\boldsymbol{x}_p))\right).
\end{align}
Note that Pannala and Menon \cite{Pannala1999} and Sankaran and Menon \cite{Sankaran2002} already applied the concept of a particle source term in the transport equation of the subgrid-scale kinetic energy, but with a different realization and without providing a rigorous derivation. In order to distinguish the present model including the particle source term Eq. \eqref{eq:ParticleSourceTerm} from the original LDKM we refer to it as modified LDKM (mLDKM). The constants $C_{\epsilon}$ and $C_{\mathrm{k}}$ are computed dynamically based on the assumption of scale similarity. The Leonard stress tensor is defined as
\begin{align}
    L_{ij} = \widehat{\tilde{u}_i\tilde{u}_j} - \hat{\tilde{u}}_i\hat{\tilde{u}}_j,
\end{align}
where $\hat{.}$ indicates a filtering operation with the filterwidth $\hat{\Delta} = 2\Delta$. Assuming that the Leonard stress tensor is analogously computed to the subgrid-scale stress tensor
\begin{align}
    L_{ij} = -2 C_{\mathrm{k}}\hat{\Delta} K_{\mathrm{test}}^{1/2} \hat{\tilde{S}}_{ij} + \dfrac{1}{3}\delta_{ij}L_{kk},
\end{align}
the constant $C_{\mathrm{k}}$ can be dynamically computed from 
\begin{align}
    C_{\mathrm{k}} = \dfrac{1}{2} \dfrac{L_{ij}\sigma_{ij}}{\sigma_{ij}\sigma_{ij}},
\end{align}
with 
\begin{align}
    \sigma_{ij} = -\hat{\Delta} K_{\mathrm{test}}^{1/2} \hat{\tilde{S}}_{ij}
\end{align}
and
\begin{align}
    K_{\mathrm{test}} = \dfrac{1}{2}(\widehat{\tilde{u}_i\tilde{u}_i} - \hat{\tilde{u}}_i\hat{\tilde{u}}_i) = \dfrac{1}{2}L_{ii}.
\end{align}
Note that $\delta_{ij}\tilde{S}_{ij}=0$ for incompressible flows. \\
Assuming the scale similarity to also be valid for the dissipation gives
\begin{align}
    C_{\epsilon}\dfrac{K_{\mathrm{test}}^{3/2}}{\hat{\Delta}} =  (\nu_{\mathrm{f}} + \nu_{\mathrm{k}}) \left( \widehat{\dfrac{\partial\tilde{u}_i}{\partial x_j} \dfrac{\partial\tilde{u}_i}{\partial x_j}} - \dfrac{\partial\hat{\tilde{u}}_i}{\partial x_j}\dfrac{\partial\hat{\tilde{u}}_i}{\partial x_j}\right).
\end{align}
With this the dynamic value of $C_{\epsilon}$ can be obtained from 
\begin{align}
    C_{\epsilon} = \dfrac{\hat{\Delta} (\nu_{\mathrm{f}} + \nu_{\mathrm{k}})}{K_{\mathrm{test}}^{3/2}} \left( \widehat{\dfrac{\partial\tilde{u}_i}{\partial x_j} \dfrac{\partial\tilde{u}_i}{\partial x_j}} - \dfrac{\partial\hat{\tilde{u}}_i}{\partial x_j}\dfrac{\partial\hat{\tilde{u}}_i}{\partial x_j}\right).
\end{align}
The original LDKM for single phase flows (without the particle source term) has some advantageous properties. The dynamical constant $C_{\mathrm{k}}$ can become negative and, thus, theoretically enables emulating the backward energy cascade. However, similar to Kim et al. \cite{Kim1997} the eddy viscosity is numerically limited to $\nu_{\mathrm{k}}>-\nu_{\mathrm{f}}$ in order to ensure a stable numerical solution of the flow equations. In contrast to the dynamic model of Germano et al. \cite{Germano1991}, no averaging along statistically homogeneous directions is required with the LDKM, which requires the existence of statistically homogeneous directions. With the dynamic computation of the constants, the model does not contain any adjustable constants. \\
Besides the theoretical advantages of the LDKM in a single phase flow, the main advantage is that it provides a framework for incorporating the particle source term in a deterministic way. If the particles remove kinetic energy from the subgrid-scales, $K_{\mathrm{sgs}}$ decreases and the fluid dissipation (i.e., the eddy viscosity) is reduced. However, the model is not capable of considering at which scales the enhancement or attenuation of turbulence takes place. In reality, the spectral distribution of the subgrid-scale kinetic energy can play an important role. \\
The particle source term $\varPhi_{\mathrm{P}}$ is not closed in the scope of LES, because the unfiltered velocity at the particle position is required, whereas only the filtered velocity is available. Thus, the mLDKM including the particle source term is only applicable if a suitable model for the subgrid-scale velocity at the particle position is provided.

\subsection{Approximate reconstruction of the subgrid-scale velocity}
\label{ssec:sgsvelocity}
The computation of the drag force acting on the particle requires the knowledge of the fluid velocity at the particle positions, which is used in the equations of motion of the particles and as the feed-back force on the fluid. In addition to the LES velocity, the subgrid-scale velocity has to be provided. We approximate the subgrid-scale velocity as a truncated Fourier-series expansion
\begin{align}
\label{eq:SeriesExpansionSGVelocity}
    \boldsymbol{u}^{\prime} = \sum_{m=0}^{N_{\mathrm{m}}-1}\left( \boldsymbol{A}_m(t) \cos(\boldsymbol{k}_m \cdot \boldsymbol{x}) + \boldsymbol{B}_m(t) \sin(\boldsymbol{k}_m \cdot \boldsymbol{x}) \right),
\end{align}
where $\boldsymbol{A}_m(t)$ and $\boldsymbol{B}_m(t)$ are time dependent coefficients, $N_{\mathrm{m}}$ the number of modes and $\boldsymbol{k}_m$ the wave vectors. With similar coefficients for the entire domain, the subgrid-scale velocity is statistically homogeneous. In order to overcome the limitation of global statistical homogeneity, statistically homogeneous sub-domains $\Omega_{\mathrm{domain}} \in \Omega$ are defined that own a distinct set of coefficients $\boldsymbol{A}_m(t)$ and $\boldsymbol{B}_m(t)$, respectively. Quantities that are known in the LES may be averaged over the sub-domain
\begin{align}
    \langle \phi \rangle_{\mathrm{domain}} = \dfrac{1}{V_{\mathrm{domain}}}  \int\displaylimits_{\Omega_{\mathrm{domain}}} \phi \mathrm{d}V,
\end{align}
where $V_{\mathrm{domain}}$ indicates the volume of a sub-domain. We exploit the findings of Lavel et al. \cite{Laval2001} that suggest that the effects of the non-linear term in the governing equations for the subgrid-scale velocity on the kinetic energy spectrum and intermittency may be replaced by an additional viscosity, that can be obtained from renormalization groups \cite{Canuto1996a}
\begin{align}
\label{eq:TurbulentViscositySmall}
    \nu_{\mathrm{t}}^{\prime}(k) =  \left(\nu^2_{\mathrm{f}} + C_{\nu} \int\displaylimits_k^{\infty}q^{-2} E(q) \mathrm{d}q \right)^{1/2} - \nu_{\mathrm{f}},
\end{align}
with the fluid kinetic energy spectrum $E$ and an analytical constant $C_{\nu}=2/5$. An equation to obtain a preliminary set of coefficients $A_{m,i}^*$ and $B_{m,i}^*$ can be derived \cite{Hausmann2022a}
\begin{align}
\label{eq:Coefficients1}
    \dfrac{A_{m,i}^* - A_{m,i}^n}{\Delta t} + \langle\tilde{u}_j^n\rangle_{\mathrm{domain}} \left( k_{m,j} B_{m,i}^n + \dfrac{\partial A_{m,i}^n}{\partial x_j}\right)+
    A_{m,j}^n \langle\dfrac{\partial \tilde{u}_i^n}{\partial x_j}\rangle_{\mathrm{domain}} = \nonumber\\ 
    (\nu_{\mathrm{f}}+\nu_\mathrm{t}^{\prime}) \left( -\lvert \boldsymbol{k}_m \rvert^2 A_{m,i}^n + \dfrac{\partial^2 A_{m,i}^n}{\partial x_j \partial x_j} + 2 k_{m,j} \dfrac{\partial B_{m,i}^n}{\partial x_j}\right) + f_{m,i},   \\
\label{eq:Coefficients2}
    \dfrac{B_{m,i}^* - B_{m,i}^n}{\Delta t} + \langle\tilde{u}_j^n \rangle_{\mathrm{domain}} \left( \dfrac{\partial B_{m,i}^n}{\partial x_j}- k_{m,j} A_{m,i}^n\right) + 
    B_{m,j}^n \langle\dfrac{\partial \tilde{u}_i^n}{\partial x_j}\rangle_{\mathrm{domain}} = \nonumber\\
    (\nu_{\mathrm{f}}+\nu_\mathrm{t}^{\prime}) \left( -\lvert \boldsymbol{k}_m \rvert^2 B_{m,i}^n  +\dfrac{\partial^2 B_{m,i}^n}{\partial x_j \partial x_j} - 2 k_{m,j}\dfrac{\partial A_{m,i}^n}{\partial x_j}\right)+ g_{m,i}.
\end{align}
Note that no summation over the index $m$ is carried out. The index $n$ indicates the time level and $f_{m,i}$ and $g_{m,i}$ are forcing terms to maintain a desired kinetic energy of the subgrid-scales. The coefficients are made divergence free by applying a subsequent projection operation to the preliminary coefficients
\begin{align}
\label{eq:Projection1}
    \boldsymbol{A}_m^{n+1}(t) = \boldsymbol{A}_m^*(t) - \boldsymbol{k}_m\dfrac{\boldsymbol{k}_m\cdot \boldsymbol{A}_m^*(t)}{\lvert \boldsymbol{k}_m \rvert^2}, \\
\label{eq:Projection2}
    \boldsymbol{B}_m^{n+1}(t) = \boldsymbol{B}_m^*(t) - \boldsymbol{k}_m\dfrac{\boldsymbol{k}_m\cdot \boldsymbol{B}_m^*(t)}{\lvert \boldsymbol{k}_m \rvert^2}.   
\end{align}
Since with discretized spatial derivatives the solution for the coefficients essentially consists of explicit algebraic operations, the numerical solution is rather inexpensive. Numerical experiments show that the solution for the coefficients $A_{m,i}$ and $B_{m,i}$ requires a computing time of the same order as the LES \cite{Hausmann2022a}.\\
The fact that every sub-domain possesses a distinct set of coefficients requires the interpolation of the coefficients between the sub-domains. In Hausmann et al. \cite{Hausmann2022a} an interpolation of the coefficients is presented, that leads to a divergence free subgrid-scale velocity field between sub-domains. 

\subsection{Coupling between the fluid-phase and the particle-phase}
\label{ssec:couplingbewteenphases}
\begin{figure}[htbp]
    %classical LES
    \begin{adjustbox}{minipage=\linewidth,scale=0.8}
    \centering
    \begin{tikzpicture}[node distance=5cm]
    \node (vf) [process] {$\boldsymbol{v}_p$}; 
    \node (ff) [process, right of=vf] {$\boldsymbol{F}_{\mathrm{D},p}$};
    \node (uf) [process, right of=ff] {$\tilde{\boldsymbol{u}}$};
    \node (sgsmod)[whiteprocess] at (5,-2.5) {$\tau_{ij}$ from single phase model (e.g., LDKM)};
 
    \draw [latex-latex, thick] (vf) to (ff);
    \draw [latex-latex, thick] (uf) to (ff);   
    \draw [latex-latex, thick] (sgsmod) to node[anchor=south, xshift=0.9cm, yshift=0.5cm] {$\tau_{ij}$}(uf);
    
    \node at (6.4,-1.5) {$\tilde{\boldsymbol{u}}$};   
    \node at (0,1.1) {\Large particle velocity};
    \node at (5,1.1) {\Large coupling force};
    \node at (10,1.1) {\Large fluid velocity};   

    \node at (-5,-1) {\Large \textbf{classical LES:}};

    %empty node for the distance between the tikz pictures
    \node at (0,-5){};
    \end{tikzpicture}
    \end{adjustbox}
    
    %modeled LES
    \begin{adjustbox}{minipage=\linewidth,scale=0.8}
    \centering
    \begin{tikzpicture}[node distance=5cm]
    \node (vf) [process] {$\boldsymbol{v}_p$}; 
    \node (ff) [process, right of=vf] {$\boldsymbol{F}_{\mathrm{D},p}$};
    \node (uf) [process, right of=ff] {$\tilde{\boldsymbol{u}}$};
    \node (mLDKM)[whiteprocess] at (5,-2.5) {$K_{\mathrm{sgs}}$ and $\tau_{ij}$ from mLDKM};
    \node (enrich)[whiteprocess, right of=mLDKM] {modeling $\boldsymbol{u}^{\prime}$};
 
    \draw [latex-latex, thick] (vf) to (ff);
    \draw [latex-latex, thick] (uf) to (ff);   
    \draw [latex-latex, thick] (mLDKM) to node[anchor=south, xshift=0.9cm, yshift=0.5cm] {$\tau_{ij}$}(uf);
    \node at (6.4,-1.5) {$\tilde{\boldsymbol{u}}$};
    \draw [-latex, thick] (ff) to (mLDKM);
    \draw [-latex, thick] (mLDKM) to node[anchor=north, xshift=0.8cm] {$K_{\mathrm{sgs}}$} (enrich);
    \draw [-latex, thick] (uf) to (enrich);
    \draw [-latex, thick] (enrich) to node[anchor=north, xshift=-1.2cm, yshift=1.2cm] {$\boldsymbol{u}^{\prime}$} (ff);
    
    \node at (0,1.1) {\Large particle velocity};
    \node at (5,1.1) {\Large coupling force};
    \node at (10,1.1) {\Large fluid velocity};   

    \node at (-5,-1) {\Large \textbf{modeled LES:}};
    \end{tikzpicture}
    \end{adjustbox}
    \caption{Visualization of the modeled interactions in a classical LES (top) compared to the modeled interactions in the modeled LES (bottom).}
    \label{fig:interactions}
\end{figure}
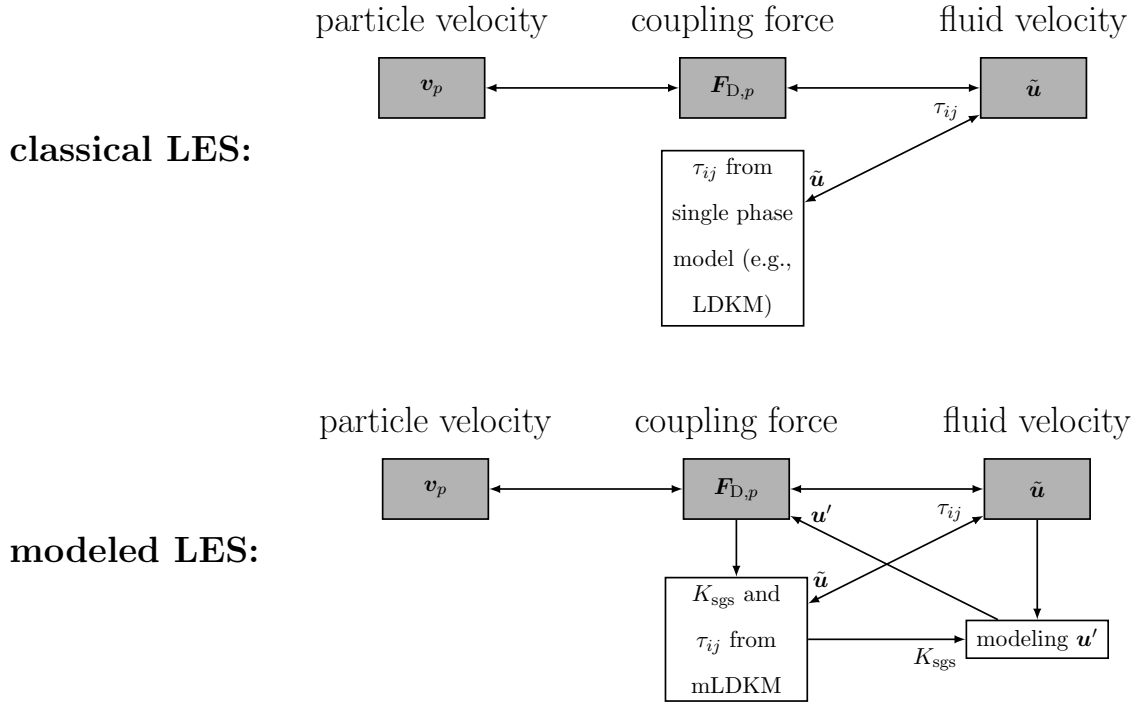

In a particle-laden LES at least three effects are not or insufficiently considered: (\romannumeral 1) The particles are accelerated by a drag force which requires the fluid velocity at the particle position. Using the filtered fluid velocity instead, leads to different clustering and Lagrangian statistics of the particles \cite{Ray2011,Fede2006a}, (\romannumeral 2) The effect of the particles on the scales that are resolved in a LES is incomplete, since the feed-back force is only computed from the filtered velocity instead of the unfiltered velocity, (\romannumeral 3) The effect of the particles on the subgrid-scales is not considered in a classical LES. The modified subgrid-scale velocity changes the subgrid-scale stress tensor compared to a single phase flow (i.e., the dissipation by the subgrid-scale fluid velocity). \\
In figure \ref{fig:interactions}, the procedure of a classical LES is compared to the newly proposed modeling framework that combines the mLDKM and the model for the subgrid-scale velocity. We refer to the latter as modeled LES. In the classical LES, the filtered fluid velocity $\tilde{\boldsymbol{u}}$ is obtained by solving the filtered NSE with a subgrid-scale stress tensor that is predicted by one of the subgrid-scale models commonly used for single phase flows. Note that subgrid-scale models for single phase flows do not incorporate information of the fluid-particle coupling force. The coupling force is computed from the filtered velocity at the particle position and is used to obtain the particle acceleration and the source terms in the fluid momentum equation. \\

The modeled LES exhibits a stronger coupling between the fluid phase and the particle phase. Using the filtered fluid velocity and the coupling force as input, the mLDKM returns a prediction of the subgrid-scale stress tensor, which is used for solving the filtered NSE and the subgrid-scale kinetic energy, which serves as target kinetic energy for the generated subgrid-scale velocity. With the modeled subgrid-scale velocity the coupling force is computed considering all turbulent length and time scales down to the Kolmogorov scales. This enables a more realistic prediction of the particle velocity, the momentum source term in the filtered NSE and the kinetic energy source term in the mLDKM. As a consequence, the subgrid-scale stress tensor and the modeled subgrid-scale velocity contain information about the turbulence modification by the particles at the unresolved scales. In theory, the modeled LES covers all occurring interactions between the fluid and the particles. \\
In section \ref{ssec:onewayresults}, we also study one-way coupling simulations with the modeled LES, in which case all arrows of figure \ref{fig:interactions} that point from the coupling force to anything else than the particle velocity vanish. The mLDKM reduces then to the original LDKM or can be replaced by any subgrid-scale model that is developed for single phase flows. The subgrid-scale kinetic energy that is required as input for the modeling of the subgrid-scale velocity can be alternatively estimated with the model of Yoshizawa \cite{Yoshizawa1986} or the dynamic model of Moin et al. \cite{Moin1991}.

\section{Simulations}
\label{sec:Simulations}
The newly proposed modeling is verified and validated by means of DNS and LES of particle-laden turbulent flows. For all the simulation cases, homogeneous isotropic turbulence is considered. Firstly, the model for the subgrid-scale velocity at the particle position is assessed by comparing the particle statistics in the modeled LES with the particle statistics in a DNS of the same one-way coupled case of statistically stationary turbulence. In a second case, the feed-back force of the particles on the fluid is investigated. This requires the full modeling framework as described in section \ref{ssec:couplingbewteenphases}, which is evaluated by comparing the modeled LES and DNS of two-way coupled particle laden decaying turbulence. The subsequent sections provide details of the realizations and configurations of the simulations.

\subsection{Solving the governing equations}
The NSE are solved with a finite volume approach that is second order in space and time \cite{Denner2020}. The continuity equation and the momentum equations are coupled using momentum weighted interpolation \cite{Bartholomew2018}. Therefore, two distinct velocity fields exist numerically, a cell centered velocity accurately satisfying the momentum balance and a face centered velocity conserving mass. \\
Statistically steady turbulence is maintained by supplying the flow with energy through source terms in the momentum equations \cite{Mallouppas2013a}. An important property of the forcing is that the artificial source terms can be introduced in a limited range of wave numbers, $k\in [k_{\mathrm{start}}, k_{\mathrm{end}}]$. For particle-laden flows, this is essential to avoid directly impacting the particle behavior by the forcing. \\
The particle equations of motion \eqref{eq:particleeom1} and \eqref{eq:particleeom2} are solved with the Verlet-scheme \cite{Verlet1967}. The drag force acting on the Lagrangian particle $p$ is computed from
\begin{align}
   \boldsymbol{F}_{\mathrm{D},p} = C_{\mathrm{D}} \dfrac{\rho_{\mathrm{f}}}{8} \pi d_p^2 |\boldsymbol{u}_{\mathrm{rel}}| \boldsymbol{u}_{\mathrm{rel}},
\end{align}
with the drag coefficient from the Schiller-Naumann correlation \cite{Schiller1933}
\begin{align}
    C_{\mathrm{D}} = \dfrac{24}{Re_p}(1+0{.}15 Re_p^{0{.}687}), 
\end{align}
and the particle Reynolds number $Re_p=|\boldsymbol{u}_{\mathrm{rel}}|d_p/\nu_{\mathrm{f}}$. The relative velocity $\boldsymbol{u}_{\mathrm{rel}}$ is defined as the difference between the fluid velocity at the particle position and the particle velocity $\boldsymbol{u}_{\mathrm{rel}}=\boldsymbol{u}(\boldsymbol{x}_p)-\boldsymbol{v}$. \\
In order to obtain the fluid velocity at the particle position, an interpolation from the Cartesian grid is required. An essential property of the interpolation scheme is that the interpolated velocity needs to be divergence free. In the present study, a divergence free interpolation from the face centered velocity (that fulfills the continuity equation with high accuracy) to the particle position is applied \cite{Toth2002}. \\
In the case of the two-way coupling simulations, the PSIC method of Crowe et al. \cite{Crowe1977} is utilized. %This approximation is valid for small particle to cell size ratios \cite{Evrard2021}. 

\subsection{Simulation setups}

\subsubsection{Single phase flow setups}
The studies in this paper consist of two different flow types, (\romannumeral 1) one-way coupling simulations of forced homogeneous isotropic turbulence, and (\romannumeral 2) two-way coupling simulations of decaying homogeneous isotropic turbulence. The computational domain of both simulation types is a cube with periodic boundary conditions in all directions and a edge length of $L$. Time quantities are given with respect to a reference time $T_{\mathrm{ref}}=L/\sqrt{2/3\langle K\rangle}$, where $\langle K\rangle$ is the average kinetic energy of the fluid. \\
The setup of the single phase flow simulations is summarized in table \ref{tab:ParametersSinglePhase}. The given values correspond to the statistically steady state of the flow (before the decay) without the particles. The two-way coupled simulations also undergo a two-way coupled forcing period to obtain a statistically steady state, before the forcing is turned off. The symbols in the table correspond to the Taylor-Reynolds number $Re_{\lambda}$, the turbulent Reynolds number based on the integral length scale $Re_{l}$, the Kolmogorov time scale $\tau_{\eta}$, the Taylor micro scale $\lambda$, and the longitudinal integral length scale $l_{11}$. \\
The DNS are carried out on a grid consisting of $N^3=256^3$ grid cells, which leads to a product of the maximum resolvable wave number $k_{\mathrm{max}}$ and the Kolmogorov length scale of $k_{\mathrm{max}} \eta = 1{.}37$. The LES are solved on a grid with $N^3=32^3$ cells. 
%T_ref = 2*pi/sqrt(2/3) = 7.70
\begin{table}[htb]
\caption{\label{tab:ParametersSinglePhase}Single phase flow parameters of the HIT simulation configuration.}
\begin{ruledtabular}
\begin{tabular}{cc}
Parameters & Values\\
\hline
$Re_{\lambda}$ & $75$ \\
$Re_{l}$ & $205$ \\
$\eta/L$ & $0{.}0017$ \\
$\tau_{\eta}/T_{\mathrm{ref}}$ & $0{.}0075$ \\
$\lambda/L$ & $0{.}029$ \\
$l_{11}/L$ & $0{.}079$ \\
$k_{\mathrm{start}}L/2\pi$ & $3$ \\
$k_{\mathrm{end}}L/2\pi$ & $6$ \\
\end{tabular}
\end{ruledtabular}
\end{table}

\subsubsection{Particle setups of the one-way coupled case}
\label{ssec:onewaycase}
In the one-way coupled case, particles of five different Stokes numbers, $St=\tau_p/\tau_{\eta}$, are introduced in the previously defined flow setup. The Stokes numbers are based on the Kolmogorov time scale of the statistically steady single phase flow. Since in the one-way coupling simulations the flow does not experience any feedback by the particles, the previously defined parameters of the flow do not change. \\
The particle relaxation time, $\tau_p=\rho_p d_p^2/(18 \rho_{\mathrm{f}} \nu_{\mathrm{f}})$, the number of particles of the respective simulation $N_{\mathrm{p}}$, and the particle diameter to mesh spacing ratio $d_p/\Delta h$ are summarized in table \ref{tab:ParametersDNS256}. \\
The simulations run more than $150 \tau_{\eta}$ before the statistics are evaluated to obtain converged statistics that are independent of the initial conditions.

\begin{table}[htb]
\caption{\label{tab:ParametersDNS256}Particle configurations of the one-way coupling case.}
\begin{ruledtabular}
\begin{tabular}{cccccc}
Parameter & St05 & St1 & St2 & St4 & St8\\
\hline
$\tau_p/T_{\mathrm{ref}}$ & $0.0037$ & $0.0075$ & $0.015$ & $0.03$ & $0.06$ \\
$N_{\mathrm{p}}$ & $480115$ & $480115$ & $480115$ & $480115$ & $480115$ \\
$d_p/\Delta h$ & $0.2$ & $0.2$ & $0.2$ & $0.2$ & $0.2$ \\
\end{tabular}
\end{ruledtabular}
\end{table}

\subsubsection{Particle setups of the two-way coupled case}
\label{ssec:twowaycase}
Two simulation configurations are performed including two-way coupled particles of two different Stokes numbers. Both configurations are started with a period of forced turbulence until the particle-laden turbulence reaches a statistically steady state (at least $150 \tau_{\eta}$). The decay of the turbulence and the tracking of the statistics starts after this forcing period. Note that the Kolmogorov time scale and thus also the Stokes numbers are based on the statistically steady single phase flow turbulence. The particle related parameters of the two two-way coupled simulations are summarized in table \ref{tab:ParametersDNS128}.
\begin{table}
\caption{\label{tab:ParametersDNS128}Particle configurations of the two-way coupling case.}
\begin{ruledtabular}
\begin{tabular}{ccc}
Parameter & St2 & St8\\
\hline
$\tau_p/T_{\mathrm{ref}}$ & $0.015$ & $0.060$ \\
$N_{\mathrm{p}}$ & $45830011$ & $12057066$ \\
$d_p/\Delta h$ & $0.1$ & $0.1$ \\
$\phi$ & $1.0$ & $1.0$  \\
\end{tabular}
\end{ruledtabular}
\end{table}

\subsubsection{Parameters of the modeling framework}
Besides the DNS and the classical LES, we also conduct simulations using the proposed LES modeling framework. For the mLDKM part of the modeling framework, there are no tunable model parameters that have to be specified. The model for the subgrid-scale velocity however, requires to choose some parameters. It was shown by Hausmann et al. \cite{Hausmann2022a}, that the statistics of the subgrid-scale velocity are relatively insensitive to the values of these parameters.  \\
The number of statistically homogeneous sub-domains, $N_{\mathrm{domain}}$ depends on the characteristic length scales at which the statistics of the subgrid-scale velocity vary. Based on experience, we suggest to choose the size of a sub-domain approximately four times the size of a LES grid cell per direction. Note that in previous studies, the number of sub-domains did not critically influence the velocity statistics. \\
Another parameter is related to the interpolation between the sub-domains in order to obtain a divergence free velocity field. The interpolation kernel is \cite{Hausmann2022a}
\begin{align}
    W(r) = 1-\dfrac{1}{1+\mathrm{e}^{-\alpha r}},
\end{align}
where $r$ is the distance to the respective sub-domain boundary, and $\alpha$ a parameter that determines the thickness of the region that is influenced by the interpolation. In general, the influence region of the interpolation should be as small as possible to keep the region that is not affected by the interpolation as large as possible. However, the influence region should not be so small, that the particles experience the sub-domain boundary as discontinuity of the subgrid-scale velocity. We empirically found that $\alpha = 40/\Delta h_{\mathrm{domain}}$ works well for the considered cases, where $\Delta h_{\mathrm{domain}}$ is the width of a sub-domain. It is shown in the Appendix \ref{sec:Appendix}, that the clustering of the particles is not significantly influenced even if the parameter $\alpha$ is varied over a wide range. \\
The remaining parameter that has to be set is the number of modes $N_m$ in the series expansion equation \eqref{eq:SeriesExpansionSGVelocity}. Similar to the number of sub-domains, it has been shown previously that the influence of $N_{\mathrm{m}}$ on the velocity statistics is negligible as long as $N_{\mathrm{m}} = \mathcal{O}(10^2)$. The particular choice for $N_{\mathrm{m}}$ in the present study is mainly based on load balancing considerations. The values of the parameters in the present study are summarized in table \ref{tab:ParametersEnrichment}.
\begin{table}[htb]
\caption{\label{tab:ParametersEnrichment}Parameters of the model for the subgrid-scale velocity.}
\begin{ruledtabular}
\begin{tabular}{cc}
Parameter & Value\\
\hline
$N_{\mathrm{domain}}$ & $8$ \\
$\alpha$ & $40/\Delta h_{\mathrm{domain}}$  \\
$N_{\mathrm{m}}$ & $108$  \\
\end{tabular}
\end{ruledtabular}
\end{table}
\section{Results and discussions}
\label{sec:Results}

\subsection{One-way coupled configurations}
\label{ssec:onewayresults}
\begin{figure}[h]
    \centering
    \includegraphics[scale=0.75]{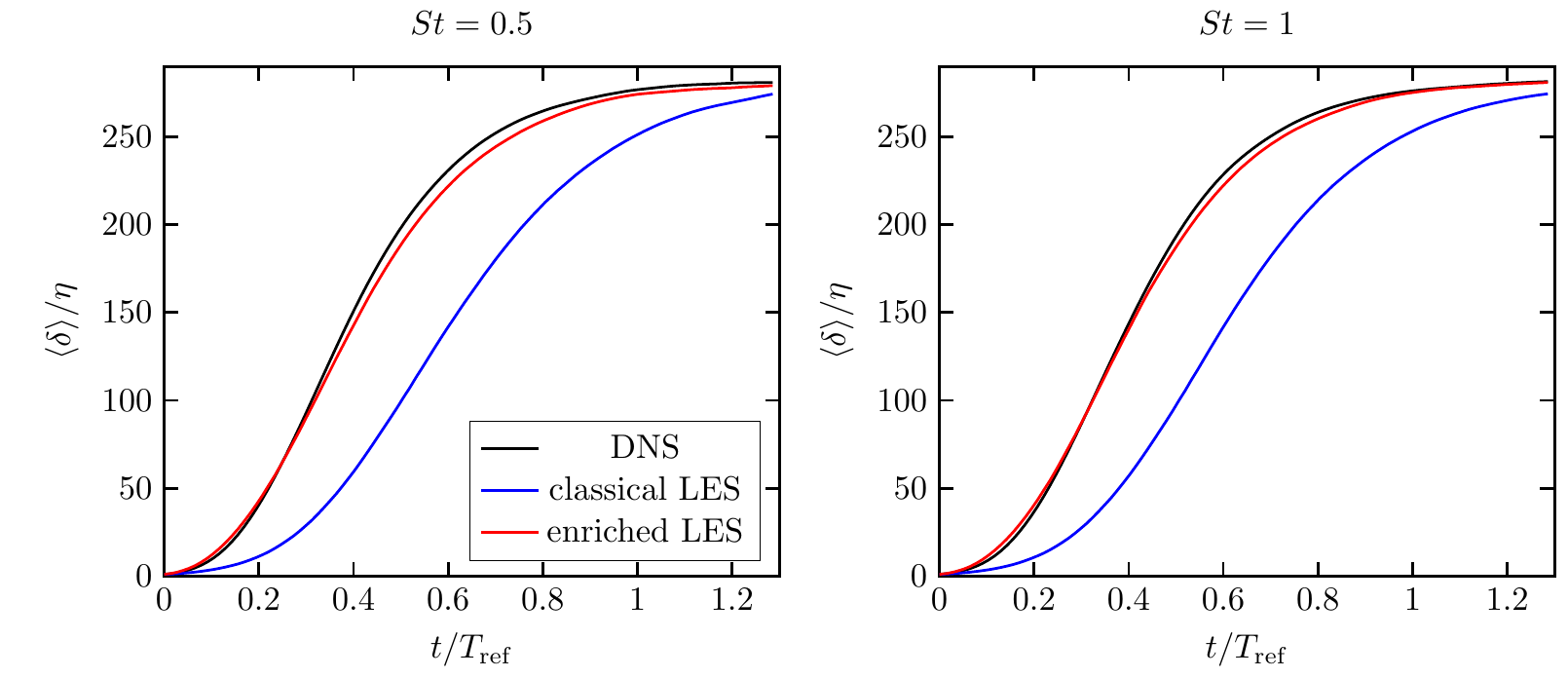}
    \includegraphics[scale=0.75]{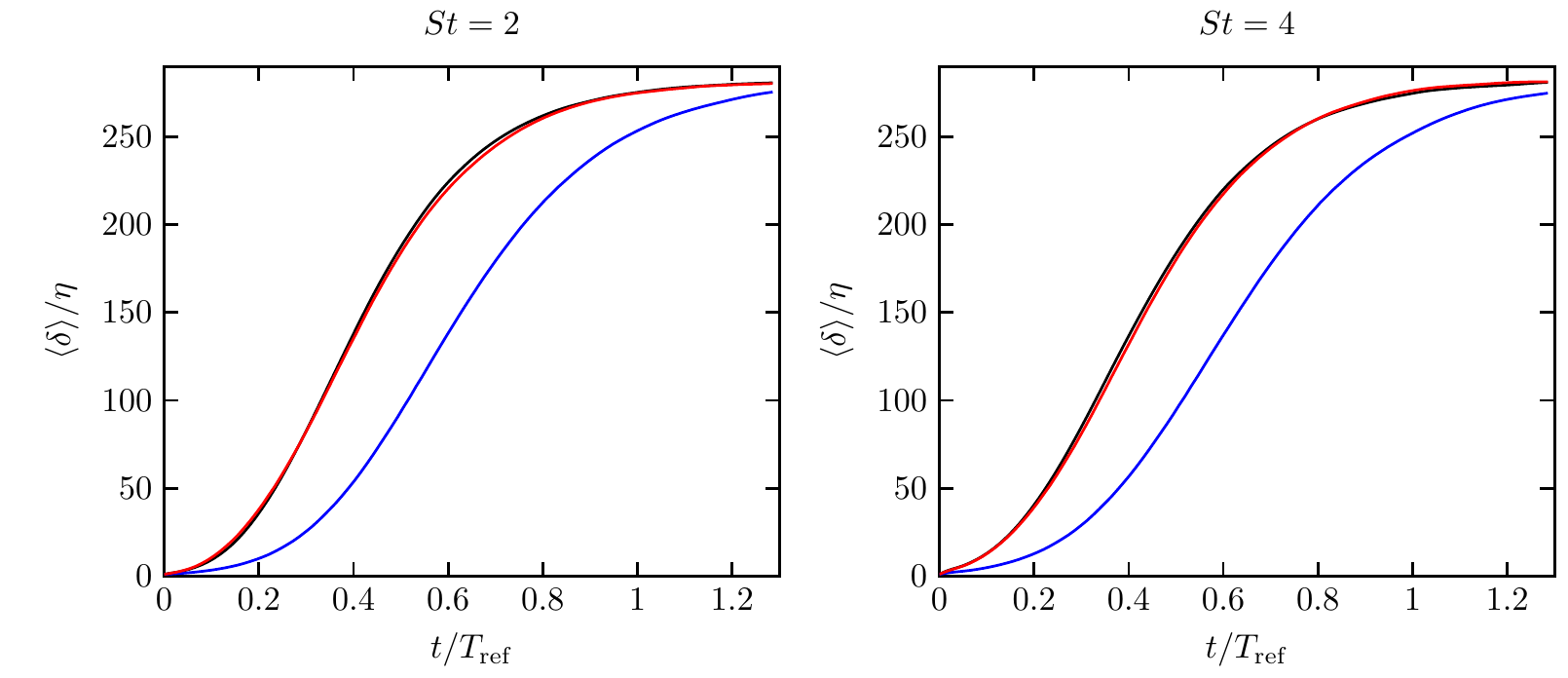}
    \includegraphics[scale=0.75]{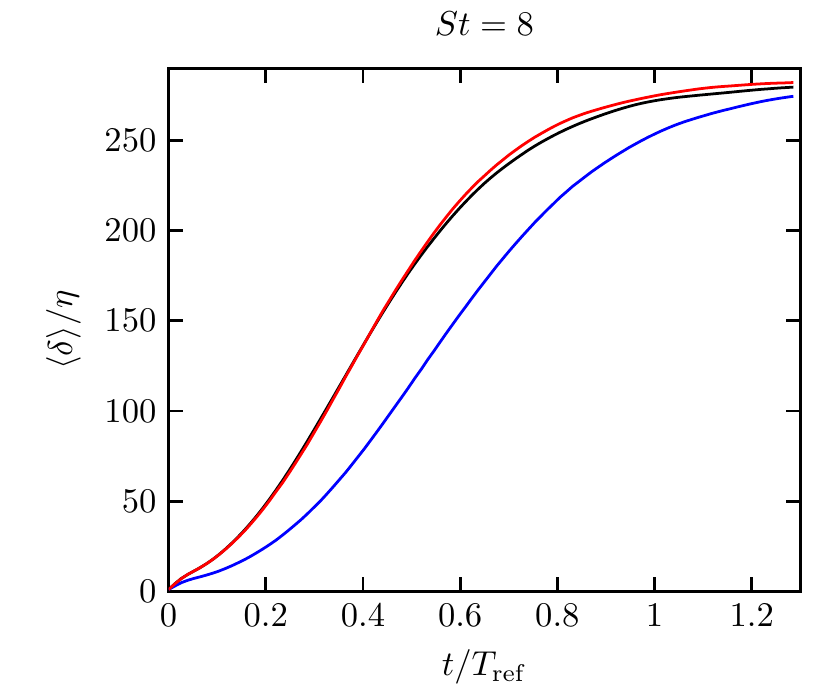}
    \caption{Particle pair dispersion of the one-way coupled simulations for different Stokes numbers in forced HIT with the flow parameters given in table \ref{tab:ParametersSinglePhase}. The results are shown for the DNS, the classical LES and the enriched LES. }
    \label{fig:pps}
\end{figure}

\begin{figure}[h]
    \centering
    \includegraphics[scale=0.75]{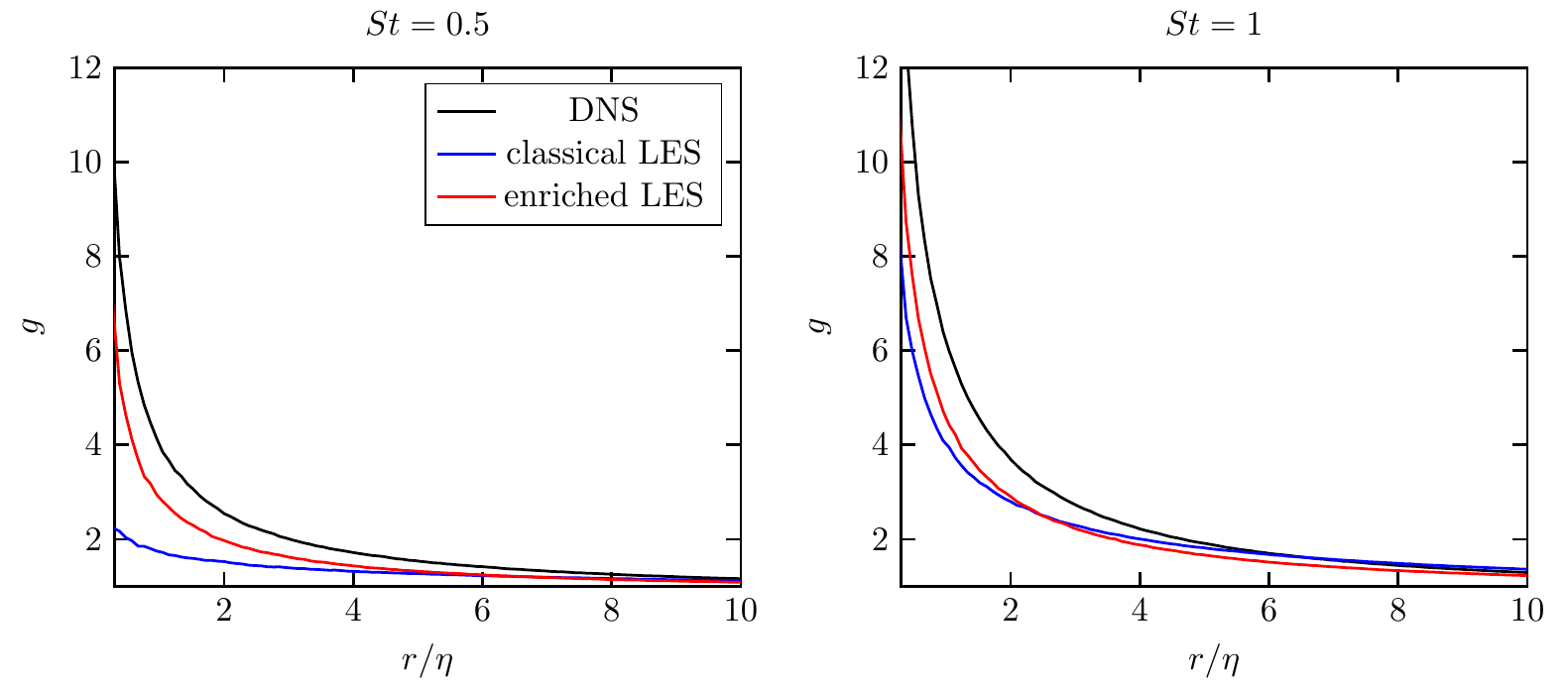}
    \includegraphics[scale=0.75]{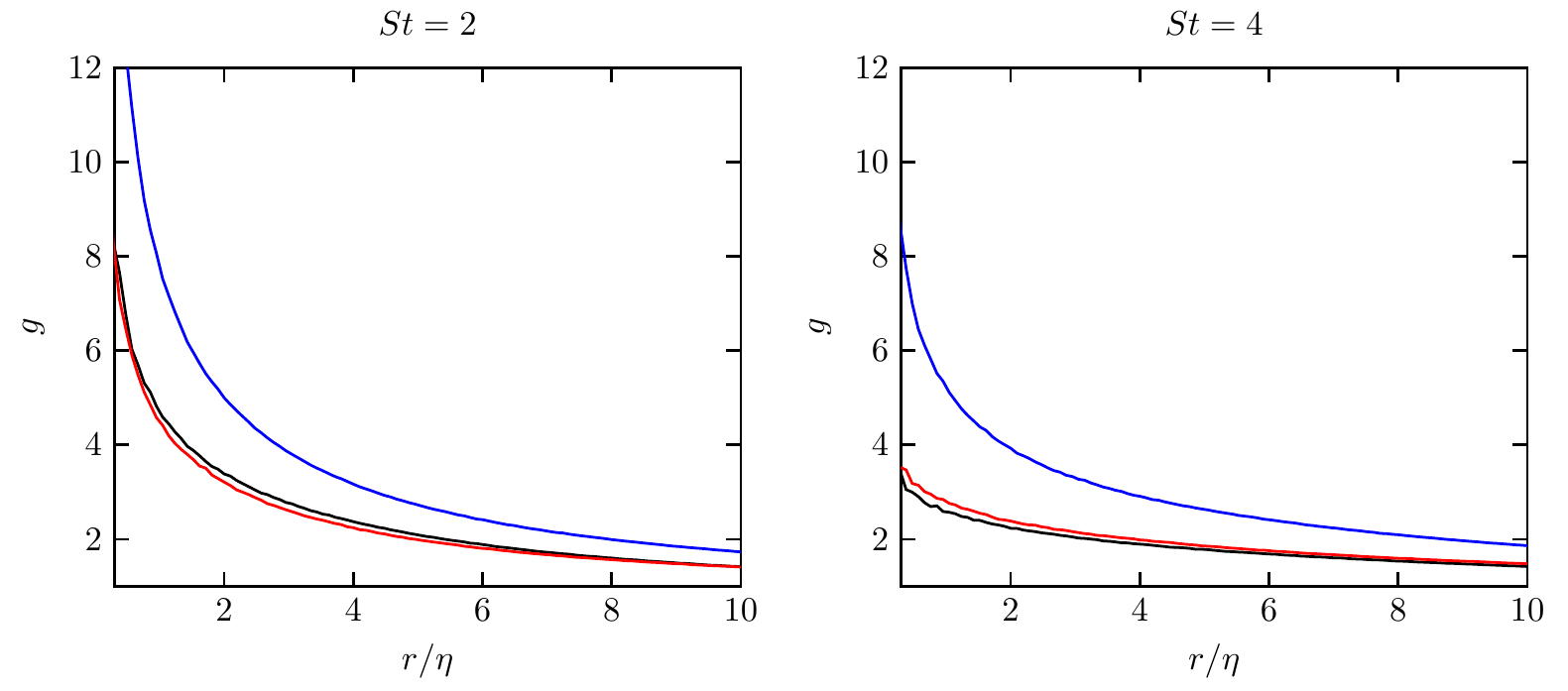}
    \includegraphics[scale=0.75]{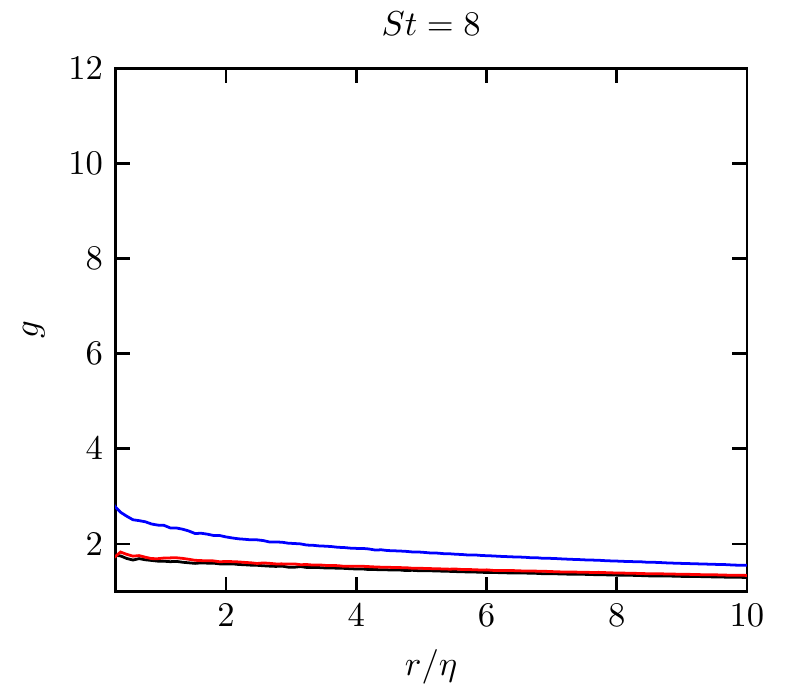}
    \caption{Radial distribution function of the one-way coupled simulations for different Stokes numbers in forced HIT with the flow parameters given in table \ref{tab:ParametersSinglePhase}. The results are shown for the DNS, the classical LES and the enriched LES.}
    \label{fig:rdf}
\end{figure}

In this section the configurations described in section \ref{ssec:onewaycase} are investigated. In the following it is referred to as enriched LES if the particles are propagated with a drag force based on the sum of the LES velocity and the modeled subgrid-scale velocity at the particle positions. \\
The particle pair dispersion is evaluated in the enriched LES, the classical LES and the DNS. The particle pair dispersion is defined as the ensemble averaged and time dependent distance between particle pairs, whereas a particle pair is considered as two particles with an initial separation of the Kolmogorov length scale
\begin{align}
    \langle \delta \rangle(t) = \langle |\boldsymbol{x}_{p0}(t) - \boldsymbol{x}_{p1}(t)| \rangle,
\end{align}
where $\boldsymbol{x}_{p0}(t)$ and $\boldsymbol{x}_{p1}(t)$ are the positions of the particles belonging to a particle pair and $\langle \delta \rangle (t=0)=\eta$. \\
The particle pair dispersion in the DNS, the classical LES and the enriched LES for the five different Stokes numbers of the one-way coupling case is depicted in figure \ref{fig:pps}. It can be observed that for all the considered Stokes numbers, the particle pairs stay together for a short time before they disperse rapidly. Eventually, the average separation reaches a steady state, which corresponds approximately to the half domain size, indicating that the maximum separation that is possible in a fully periodic cubic domain is reached. For higher Stokes numbers, the particle pairs stay close to each other for a shorter time, as particles with a large Stokes numbers are more likely to have different velocities caused by their high inertia. The classical LES predicts the particle pairs to disperse much slower than in the DNS, due to the missing effect of the subgrid-scale velocity sweeping the particles into regions of different large scale velocity. This effect is observed for all investigated Stokes numbers, but is slightly more dominant for the larger Stokes numbers. \\
In the enriched LES, the predicted particle pair dispersion almost overlaps with those of the DNS. An important reason why the enriched LES performs so well is that the subgrid-scale velocity at the particle position is computed from a spatially continuous velocity field. As a consequence, two particles that are very close also experience a similar subgrid-scale velocity. This is not the case for Lagrangian models (see, e.g. \cite{Fede2006,Bini2007,Pozorski2009}), which typically solve an evolution equation for each particle individually. Lagrangian models typically perform poorly in particle pair dispersion \cite{Marchioli2017}. \\
A property that is of high practical importance is the extent with which particles form clusters. Particle clustering can be quantified by the radial distribution function, defined as 
\begin{align}
    g(r) = \left\langle \dfrac{N_{\mathrm{p},i}(r)/\Delta V_i(r)}{N_{\mathrm{p}}/V} \right\rangle,
\end{align}
where $N_{\mathrm{p},i}(r)$ is the number of particles in a spherical shell with radius $r$ centered a the location of the original particle and $V_i(r)$ is the volume of this spherical shell. The radial distribution function is normalized by the total number of particles $N_{\mathrm{p}}$ and the total volume of the domain $V$. Values of $ g  > 1$ indicate particle clustering and $ g  = 1$ uniformly distributed particles. \\
In figure \ref{fig:rdf}, the radial distribution functions are shown. The clustering reaches its maximum at $St \approx 1$. For smaller and larger Stokes numbers, clustering is reduced. The classical LES yields an under-estimation of the clustering for $St=0{.}5$ and $St=1$ and to an over-estimation of the clustering for the Stokes numbers $St=2$, $St=4$ and $St=8$. This phenomenon has also been observed in previous studies \cite{Ray2011,Zhou2020b,Bassenne2019}. This means that the modeled subgrid-scale velocity has to increase the clustering for the small $St$ and increase the dispersion for the larger $St$. Note that increasing the particle dispersion is much simpler to achieve than increasing the particle clustering. In fact, the relations between strain and rotation of the velocity field are crucial for the correct prediction of particle clustering \cite{Maxey1987,Brandt2022}. \\
For the Stokes numbers $St=2$, $St=4$ and $St=8$, the enriched LES leads to particle clustering that almost coincides with the clustering predicted by the DNS. Even for the small Stokes numbers, the enriched LES improves the prediction of particle clustering. The clustering is significantly increased for $St=0{.}5$ but not in perfect agreement with the DNS. For the particles with Stokes number $St=1$ the enriched LES yields a good agreement with the DNS. \\
Our previous study \cite{Hausmann2022a} has shown that the probability distribution function (PDF) of the second invariant of the velocity gradient tensor is significantly improved with the enriched LES compared to the classical LES, which explains the ability of the model to increase the clustering of particles with small Stokes numbers. \\
\begin{figure}[h]
    \centering
    \includegraphics[scale=0.75]{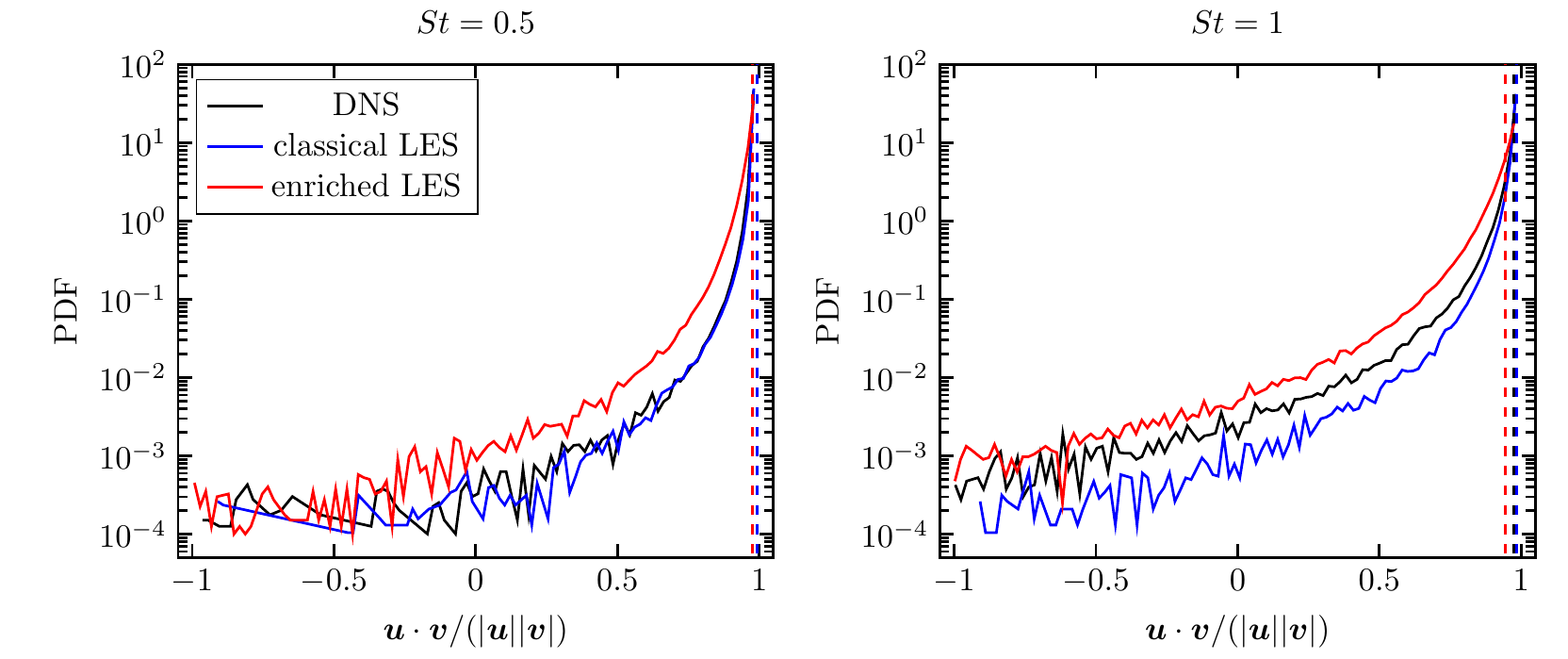}
    \includegraphics[scale=0.75]{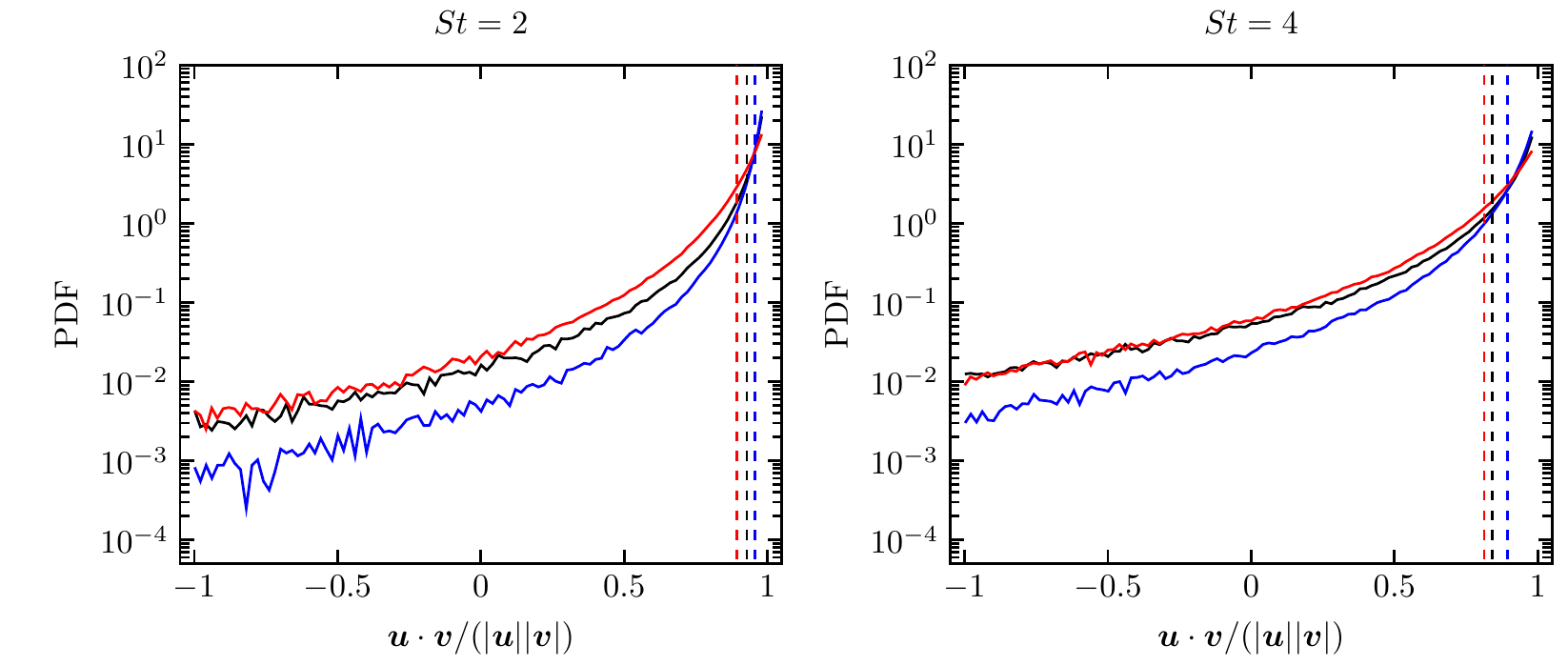}
    \includegraphics[scale=0.75]{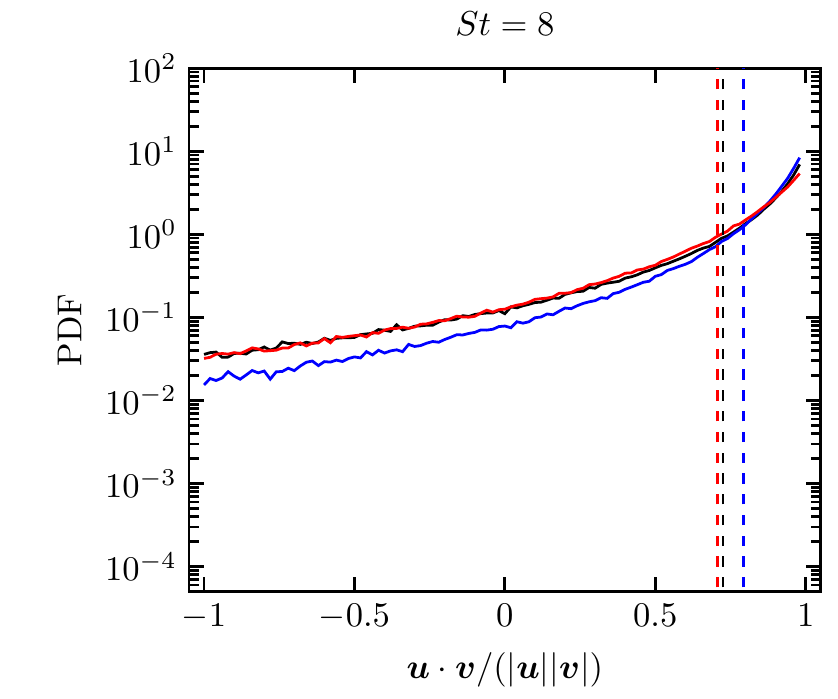}
    \caption{PDF of cosine of angle between the fluid velocity at the particle position and the particle velocity of the one-way coupled simulations for different Stokes numbers in forced HIT with the flow parameters given in table \ref{tab:ParametersSinglePhase}. The results are shown for the DNS, the classical LES and the enriched LES. The vertical lines indicate the respective mean value.}
    \label{fig:angleuv}
\end{figure}
The kinetic energy of the classical LES relative to the kinetic energy of the DNS is $K_{\mathrm{LES}}/K_{\mathrm{DNS}}\approx 0{.}83$. Together with the estimated subgrid-scale kinetic energy this gives $(K_{\mathrm{LES}}+K_{\mathrm{sgs}})/K_{\mathrm{DNS}}\approx 1{.}07$. The fact that the total kinetic energy of the enriched LES over-predicts the kinetic energy of the DNS is important for the interpretation of the following results. \\
Figure \ref{fig:angleuv} shows the PDF of the cosine of the angles between the particle velocity and the fluid velocity at the particle position, together with the mean values. The most likely event for all the considered Stokes numbers is the case where the fluid velocity is aligned with the particle velocity. However, for increasing Stokes numbers the probability of larger angles between the fluid velocity and the particle velocity also increases. Except for $St=0{.}5$, the classical LES always predicts too strong an alignment of fluid and particle velocities. For higher Stokes numbers, in particular, the particles are too heavy to follow the subgrid-scale velocity fluctuations, which typically change with high frequency and small magnitude. In the classical LES, these fluctuations are missing. The enriched LES provides the subgrid-scale fluctuations, which explains the improved PDF and means of the alignment of the velocities. For the Stokes numbers $St=0{.}5$ and $St=1$, the enriched LES over-predicts the deviation of the angle of the fluid velocity and the particle velocity, such that the results of the classical LES are closer to the DNS. However, the overall deviations from the DNS results are very minor for the small Stokes numbers. One possible reason for the deviation for $St=0{.}5$ and $St=1$, is that it is caused by the too high kinetic energy of the enriched LES. Therefore, the particles with the small Stokes numbers can not follow the velocity fluctuations as well as in the case of the DNS and the classical LES. For higher Stokes numbers, the deviation between the velocities is more significant and the higher subgrid-scale kinetic energy has a smaller influence. \\
From the results of the one-way coupled simulations it can be concluded that the enrichment with subgrid-scale velocity can significantly improve the particle statistics in HIT for a wide range of Stokes numbers. Particularly worth highlighting is that the clustering can be improved for both qualitatively different cases of $St\leq 1$ and $St>1$ with the modeled subgrid-scale velocity. 

\subsection{Two-way coupled configurations}
\begin{figure}[h]
    \centering
    \subfigure[]{\label{fig:particlesa}\includegraphics[scale=0.33]{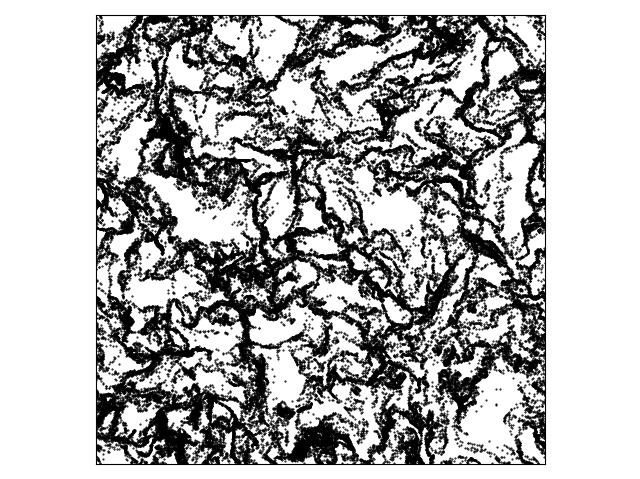}}
    \subfigure[]{\label{fig:particlesb}\includegraphics[scale=0.33]{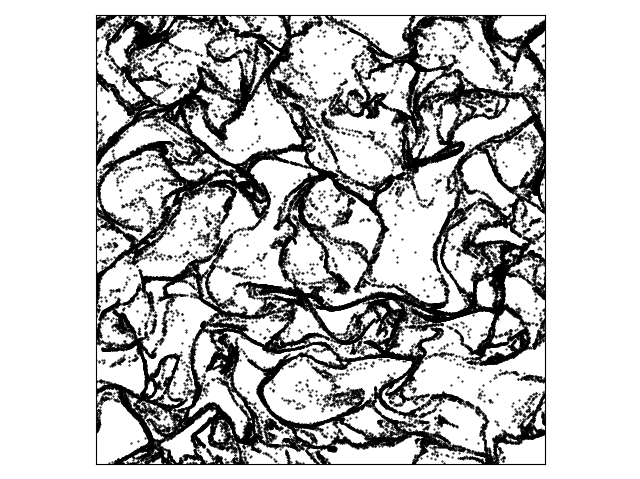}}
    \subfigure[]{\label{fig:particlesc}\includegraphics[scale=0.33]{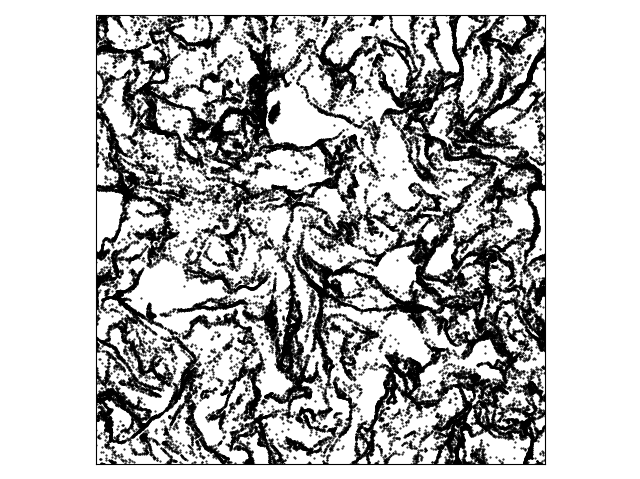}}
    \caption{Particle positions of the two-way coupled simulations with $St=2$ of HIT with the flow parameters given in table \ref{tab:ParametersSinglePhase}. The results are evaluated at the last forced time step. The particle positions are projected from a slice of thickness equal to the Kolmogorov length scale $\eta$. (a) shows the particles of the DNS, (b) the particles of the classical LES and (c) the particles of the modeled LES.}
    \label{fig:particles}
\end{figure}

In the present section, the modification of the turbulent flow by the particles is considered, which enables the assessment of the full modeling framework including the modeling of the subgrid-scale velocity combined with the mLDKM (previously introduced as modeled LES).\\
In figure \ref{fig:particles}, the projected particles in a slice of the thickness $\eta$ are plotted for the case of $St=2$ for the DNS, the classical LES, and the modeled LES. The positions are evaluated at the end of the forcing period. The particles reached steady statistics at this point in time. \\
The shapes of the clusters that are formed by the DNS and the classical LES differ significantly. The clusters of the classical LES are much coarser and more pronounced than the clusters of the DNS. The additional subgrid-scale velocity of the modeled LES breaks up the large clusters of the classical LES into clusters of smaller size, which are also less dense. \\

\begin{figure}[h]
    \centering
    \subfigure[]{\label{fig:spectraa}\includegraphics[scale=0.75]{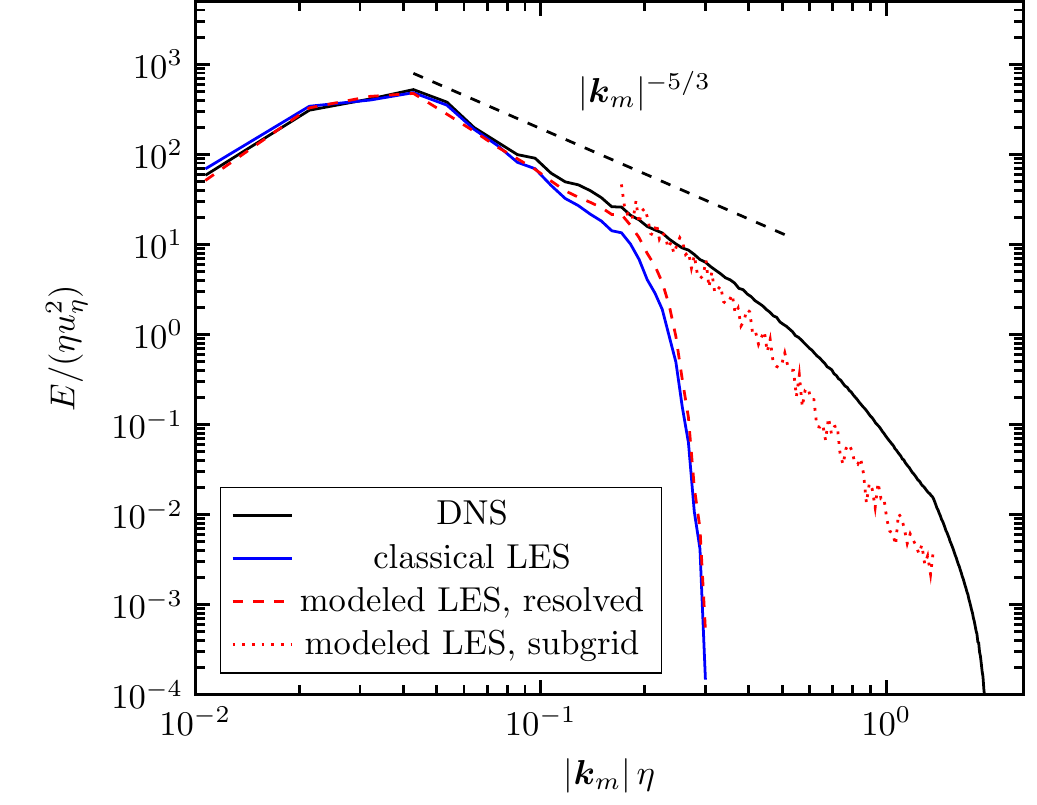}}
    \subfigure[]{\label{fig:spectrab}\includegraphics[scale=0.75]{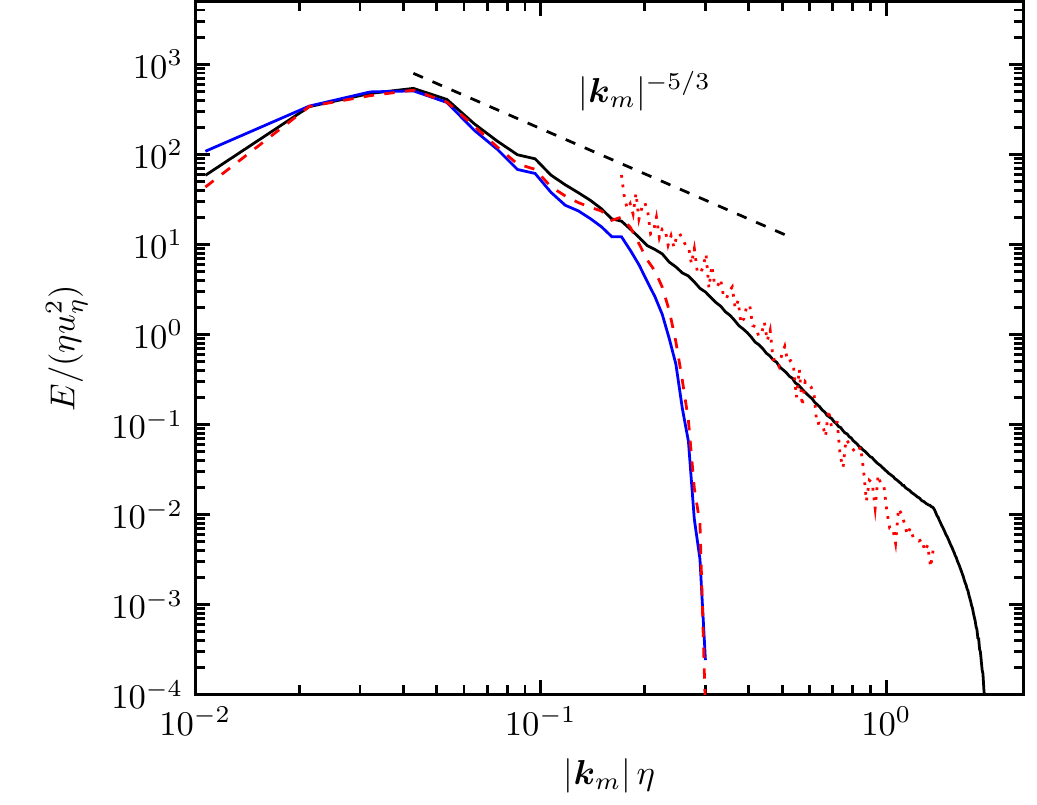}}
    \caption{Kinetic energy spectrum of the two-way coupled simulations with $St=2$ (a) and $St=8$ (b) of HIT with the flow parameters given in table \ref{tab:ParametersSinglePhase}. The results are evaluated at the last forced time step. Compared are the DNS, the classical LES and the modeled LES. The spectrum of the modeled LES is split into a part that is resolved by the LES grid and a subgrid-scale contribution that is modeled.}
    \label{fig:spectra}
\end{figure}

Figure \ref{fig:spectra} shows the kinetic energy spectra of the two-way coupled simulations for $St=2$ and $St=8$, respectively. Due to the interaction with the particles, the results obtained with the DNS deviate significantly from the inertial range slope of single phase turbulence. In addition to the DNS, the results of the classical LES and the modeled LES are depicted. The kinetic energy spectrum that is resolved by the LES grid is shown separately from the kinetic energy spectrum of the modeled subgrid-scale velocity. The classical LES over-estimates the subgrid-scale fluid dissipation, which yields a significant deviation from the DNS spectrum close to the cutoff wave number. Since the modeled LES accounts for the reduced subgrid-scale fluid dissipation due to the particle dissipation, it leads to a spectrum that is closer to the DNS spectrum for both Stokes numbers. \\
The kinetic energy spectrum of the modeled subgrid-scale velocity is in good agreement with the DNS spectrum but its shape deviates in both cases. The shape of the modeled subgrid-scale kinetic energy spectrum is very similar for $St=2$ and $St=8$ even though the shapes of the two DNS spectra are very different. The model for the subgrid-scale velocity does not receive any information on the presence of the particle, except for the kinetic energy. Therefore, the modeled subgrid-scale kinetic energy spectrum is a shifted spectrum that matches very well with a single phase flow spectrum \cite{Hausmann2022a}. \\

\begin{figure}[h]
    \centering
    \subfigure[]{\label{fig:PDFEnergyTranfera}\includegraphics[scale=0.75]{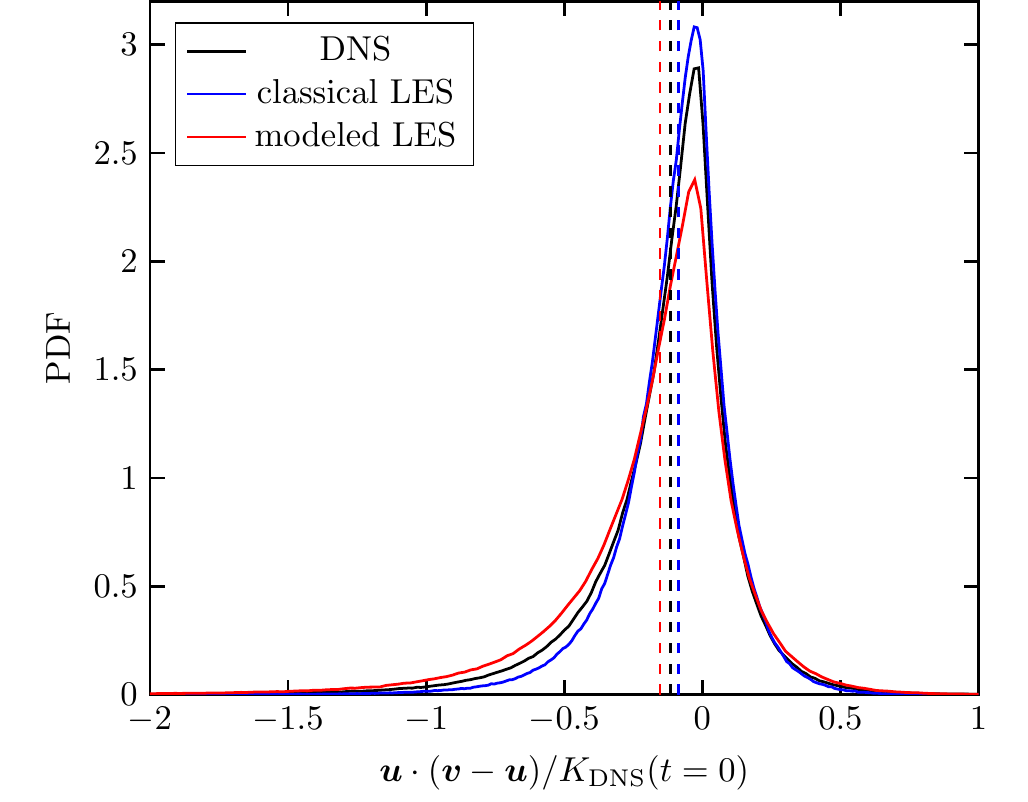}}
    \subfigure[]{\label{fig:PDFEnergyTranferb}\includegraphics[scale=0.75]{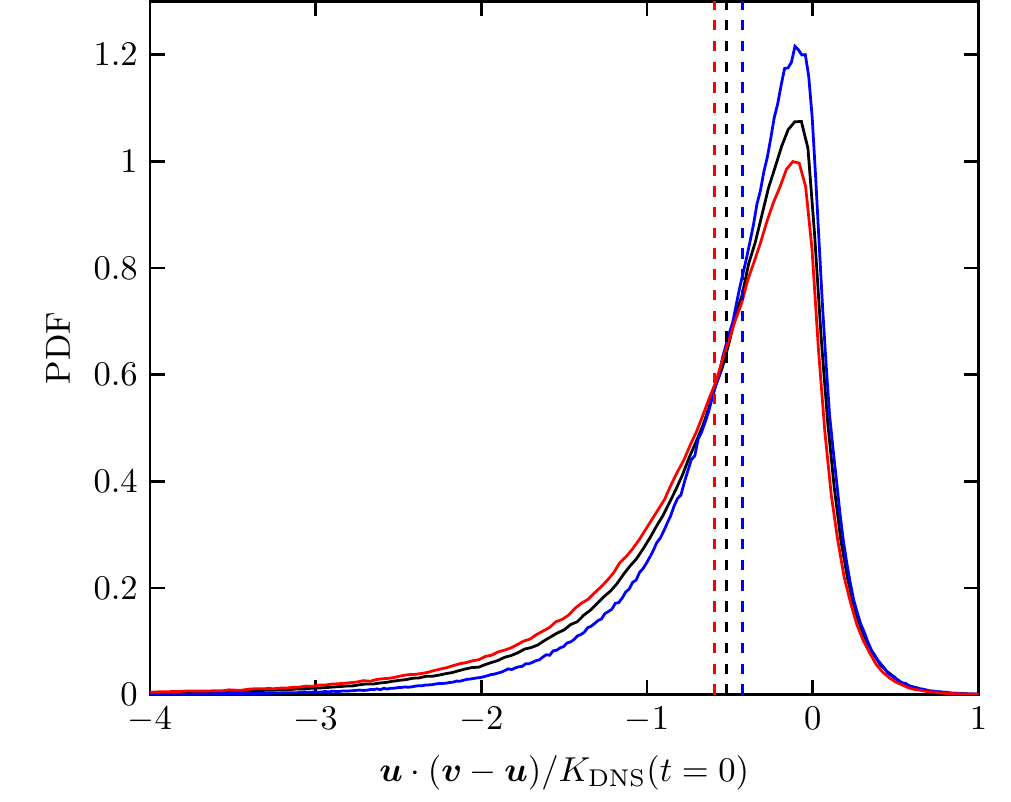}}
    \caption{PDF of the kinetic energy transfer between fluid and particles of the two-way coupled simulations with $St=2$ (a) and $St=8$ (b) of HIT with the flow parameters given in table \ref{tab:ParametersSinglePhase}. The results are evaluated at the last forced time step. The vertical lines indicate the respective mean values.}
    \label{fig:PDFEnergyTranfer}
\end{figure}

A classical LES does not fully consider the inter-phase kinetic energy transfer, since the interactions of the subgrid-scale velocity with the particles, and the unresolved particle motion with the resolved flow scales are neglected. For small particle Reynolds numbers, the kinetic energy transfer is proportional to the fluid velocity times the Stokes drag, $\boldsymbol{u}\cdot (\boldsymbol{v}-\boldsymbol{u})/\tau_p$, which is plotted in figure \ref{fig:PDFEnergyTranfer} for the present cases. Negative values indicate that kinetic energy is removed from the fluid and positive values correspond to kinetic energy that is added to the fluid by the particles. Note that, as shown by Xu et al. \cite{Xu2007a}, the energy that is removed from the fluid does not necessarily equal the kinetic energy that is added to the particles in a point-particle simulation. \\
The PDF in figure \ref{fig:PDFEnergyTranfer} shows a negative mean value for both Stokes numbers, indicating that on average the particles remove energy from the fluid. The PDF of $St=8$ is wider than the distribution of $St=2$ and possesses a larger absolute mean value. The classical LES under-predicts the absolute of the mean values of the DNS for both Stokes numbers. With the proposed modeled LES the kinetic energy that is removed by the particles is increased, which is qualitatively similar to the behavior of the DNS relative to the classical LES. In the classical LES, the particle velocities tend to align more with the local fluid velocity compared to the DNS because of the absence of small vortices that the particles can not follow. With the proposed modeling small velocity structures are provided, which increases the absolute energy transfer. However, for both Stokes numbers the width of the PDFs is slightly over-predicted by the modeled LES. \\

\begin{figure}[h]
    \centering
    \subfigure[]{\label{fig:sgsKEa}\includegraphics[scale=0.75]{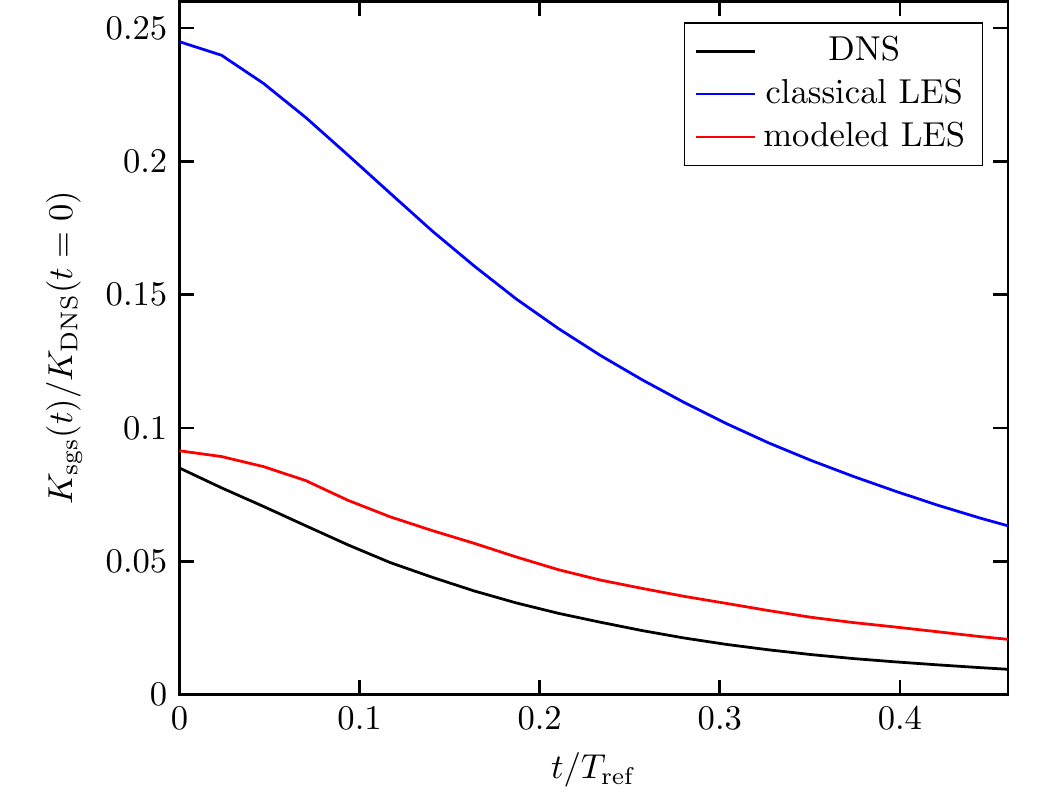}}
    \subfigure[]{\label{fig:sgsKEb}\includegraphics[scale=0.75]{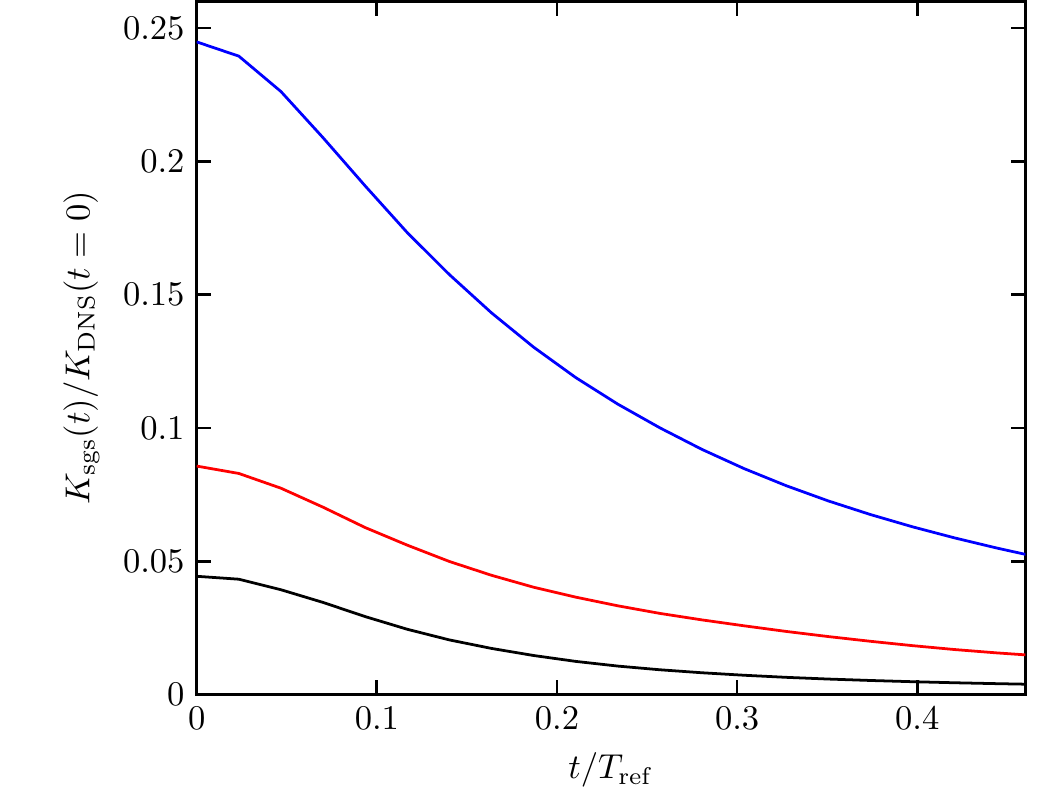}}
    \caption{Normalized subgrid-scale kinetic energy over time of the two-way coupled simulations with $St=2$ (a) and $St=8$ (b) of decaying HIT with the flow parameters given in table \ref{tab:ParametersSinglePhase}. Compared are the results of the DNS, the classical LES and the modeled LES with particle source term.}
    \label{fig:sgsKE}
\end{figure}

The second effect that is neglected in a classical LES is the reduced subgrid-scale kinetic energy due to the turbulence modification by the particles. The subgrid-scale kinetic energies over time are depicted in figure \ref{fig:sgsKE}. The subgrid-scale kinetic energy of the DNS is computed by subtracting the kinetic energy of the spectrally sharp filtered DNS from the kinetic energy of the DNS. The kinetic energies of the LES are obtained from the transport equation in the LDKM and mLKDM, respectively. It is observed that for both Stokes numbers, the classical LES predicts a much higher subgrid-scale kinetic energy than the DNS. The subgrid-scale kinetic energy of the modeled LES is significantly smaller because of the source term $\varPhi_{\mathrm{P}}$ in the transport equation for the subgrid-scale kinetic energy. A higher subgrid-scale kinetic energy yields a higher eddy viscosity and thus more subgrid-scale dissipation. The classical LES over-predicts the kinetic energy of the subgrid-scale velocity because the particle dissipation at high wave numbers is not accounted for. The additional source term in the subgrid-scale kinetic energy equation of the mLDKM considers the effect of the particles on the subgrid-scale quantities, which explains the improved results. \\

\begin{figure}[h]
    \centering
    \subfigure[]{\label{fig:decayTwoWaya}\includegraphics[scale=0.75]{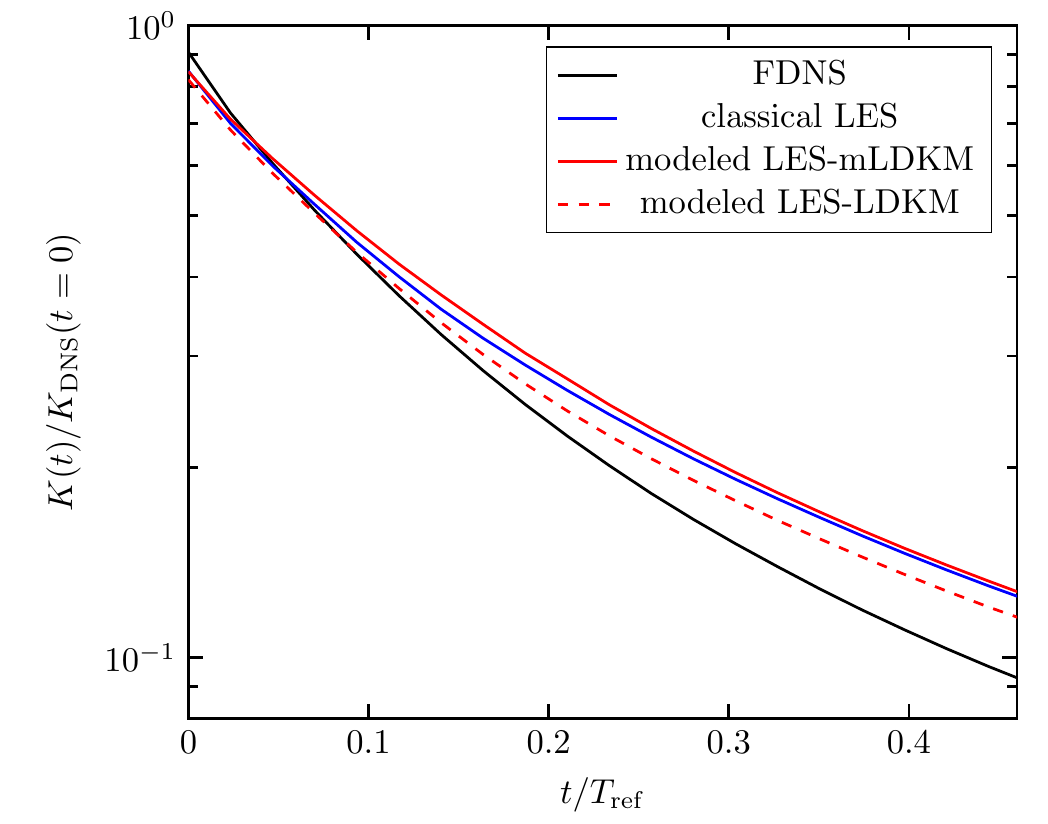}}
    \subfigure[]{\label{fig:decayTwoWayb}\includegraphics[scale=0.75]{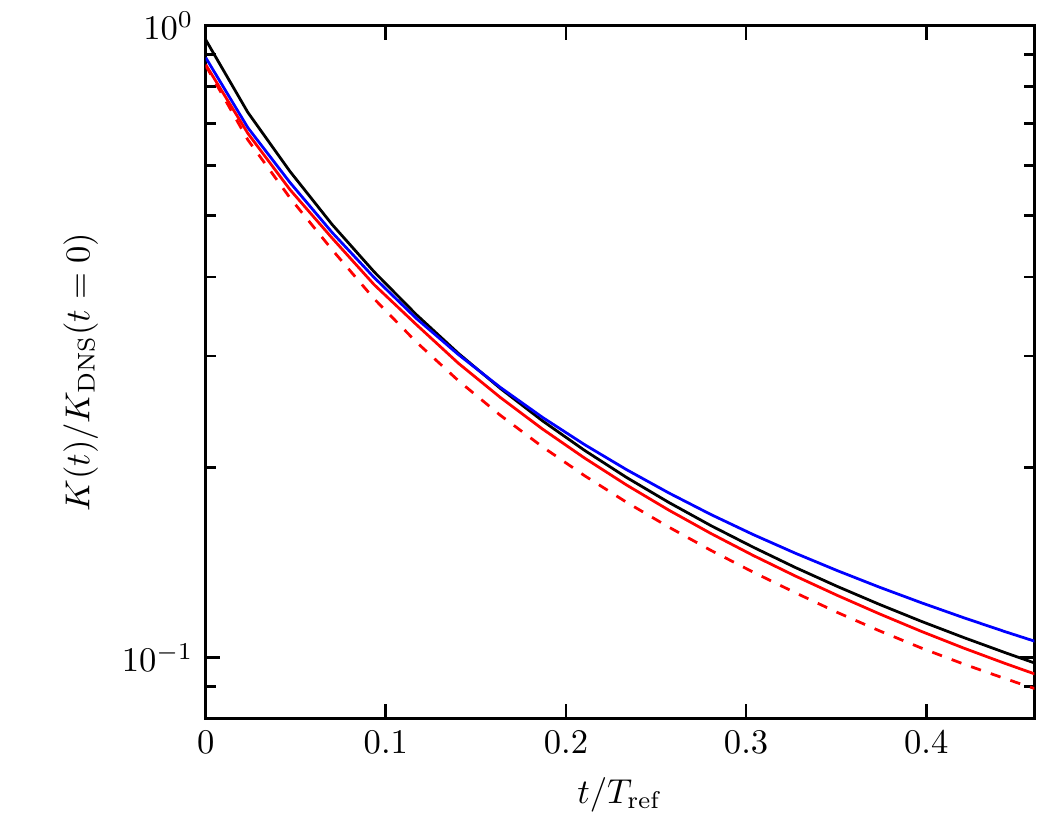}}
    \caption{Normalized kinetic energy over time of the two-way coupled simulations with $St=2$ (a) and $St=8$ (b) of decaying HIT with the flow parameters given in table \ref{tab:ParametersSinglePhase}. Compared are the FDNS, the classical LES (LES-LDKM), the modeled LES with particle source term (modeled LES-mLDKM) and the modeled LES without particle source term (modeled LES-LDKM).}
    \label{fig:decayTwoWay}
\end{figure}

It is observed that the effects that are neglected in a classical LES act in opposite directions for the considered configurations. While the classical LES under-predicts the dissipation by the particles, it over-predicts the dissipation of the subgrid-scale velocity. Thus, these two errors at least partially compensate each other, which in total may lead to fair agreement with the total kinetic energy of the DNS. However, the proposed modeled LES considers each effect (i.e., the increased particle dissipation and the reduced fluid dissipation) separately and does not rely on compensation of errors. \\
Figure \ref{fig:decayTwoWay} shows the normalized kinetic energy over time for the two-way coupled simulations with $St=2$ and $St=8$, respectively. Besides the spectrally sharp filtered DNS (FDNS) and the classical LES, the results of the modeled LES are plotted. In order to investigate the influence of the particle source term $\varPhi_{\mathrm{P}}$ in the transport equation \eqref{eq:modeledKsgs} of the subgrid-scale kinetic energy, the modeled LES is shown with the source term (modeled LES-mLDKM) and without the source term (modeled LES-LDKM). \\
For $St=2$ the LES, the modeled LES-mLDKM and the modeled LES-LDKM predict a slower decay of the fluid kinetic energy than the DNS. The deviation of the three LES cases from the DNS is much smaller for $St=8$. Both Stokes numbers show only relatively small deviations between the LES cases. However, the kinetic energy of the modeled LES-LDKM is always smaller than the kinetic energy of the modeled LES-mLDKM.\\

It is known from the literature that for Stokes numbers that are not significantly smaller than one, the total dissipation in a particle-laden flow is increased \cite{Ferrante2003,Nabavi2022,Letournel2020,Mallouppas2017}, which also applies to the Stokes numbers investigated in the present study. The total dissipation has contributions from the particles and the fluid, as it occurs in a single-phase flow. In a LES the particle and the fluid dissipation have to be modeled. As already pointed out, both contributions have opposite signs in the present cases. This becomes evident by the fact that the modeled LES without the particle source term $\varPhi_{\mathrm{P}}$ increases the total dissipation. The difference to the classical LES in this case is, that the two-way coupling force is computed using the total fluid velocity consisting of the LES velocity and the modeled subgrid-scale velocity, which leads to increased dissipation relative to the classical LES. The dissipation by the particles yields a negative source term $\varPhi_{\mathrm{P}}$, which reduces the subgrid-scale kinetic energy. This is why the total dissipation of the modeled LES-mLDKM is smaller than the total dissipation of the modeled LES-LDKM. \\
Note that both effects, the increased dissipation due to the particles and the reduced fluid dissipation, are at least qualitatively in agreement with the literature and desired. Since both effects act in opposite directions, the classical LES is still in acceptable agreement with the DNS even though it accounts for neither of the two effects. \\

\section{Conclusions}
\label{sec:Conclusions}
In the present paper, we propose a novel model for predicting the behavior of two-way coupled particle-laden flow in the framework of LES. The model accounts for the interactions that are not captured by a classical LES of a particle-laden flow, which are: (\romannumeral 1) the prediction of the particle motion due to the missing subgrid-scale fluid velocity, (\romannumeral 2) the effect of the particles on the resolved flow scales, and (\romannumeral 3) the effect of the particles on the subgrid-scales. \\
The proposed modeling framework consists of two components, a modeled transport equation for the subgrid-scale kinetic energy, that includes a source term which accounts for the modification of the subgrid-scale kinetic energy by the particles and a model for the subgrid-scale velocity, which is used to close the particle equations of motion and the source-term in the transport equation for the subgrid-scale kinetic energy. The two model components are further coupled by directly using the resulting subgrid-scale kinetic energy of the transport equation as input for the model for the subgrid-scale velocity that thus also accounts for the turbulence modulation of the subgrid-scales by the particles. \\
One-way coupled simulations are performed that are used to assess the isolated effect of missing subgrid-scale velocity in the computation of the forces acting on the particles and its modeling using the enriched LES. With the modeled subgrid-scale velocity, the particle pair-dispersion and the preferential concentration are predicted with high accuracy for a wide range of Stokes numbers while still maintaining computational costs that are reasonable for a LES. Even for the intricate case of a small Stokes number, the clustering is significantly improved. Furthermore, two-way coupled simulations of decaying HIT are carried out that require modeling of the turbulence modulation by the particles. The coupled framework yields an increased particle dissipation compared to the classical LES by considering the modeled subgrid-scale velocity in the feed-back force. The subgrid-scale fluid dissipation is decreased relative to the classical LES because the mLDKM predicts a subgrid-scale kinetic energy that considers the turbulence modulation by the particles. Both effects are in agreement with the observed physics in a DNS. As a consequence, we observe a kinetic energy spectrum with the newly proposed modeling, that is in very good agreement with the spectrum observed in the DNS. \\
Finally, it is important to mention that the proposed modeling strategy possesses the prerequisites for simulating inhomogeneous and anisotropic flows since the subgrid-scale enrichment is formulated on a grid of statistically homogeneous sub-domains,  which allows for spatially varying statistics. Considering this, the present modeling framework has the potential to improve the capabilities of LES of particle-laden turbulent flows for a wide range of applications.

\begin{acknowledgments}
This research was funded by the Deutsche Forschungsgemeinschaft (DFG, German Research Foundation) \textemdash Project-ID 457509672. 
\end{acknowledgments}

\section*{Appendix}
\label{sec:Appendix}
\begin{figure}[h]
    \centering
    \includegraphics[scale=0.75]{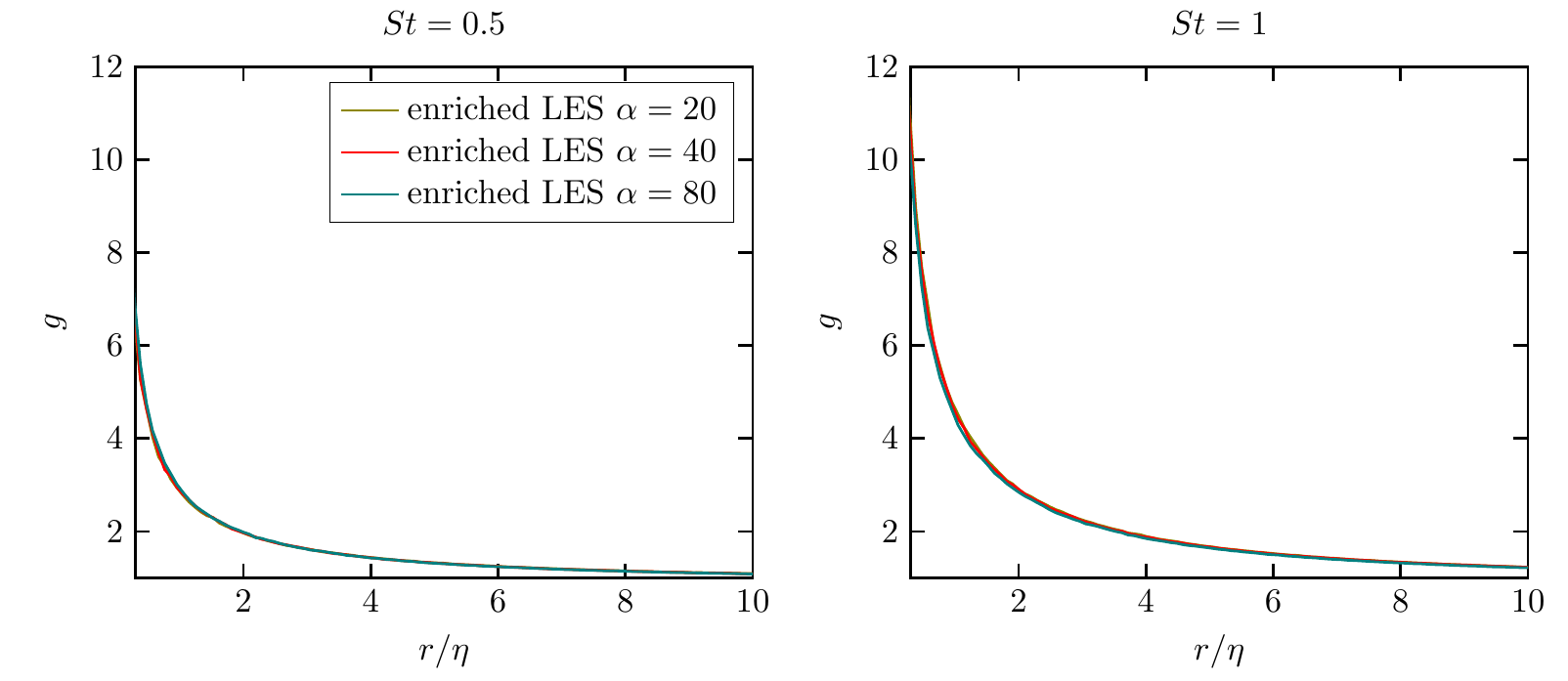}
    \includegraphics[scale=0.75]{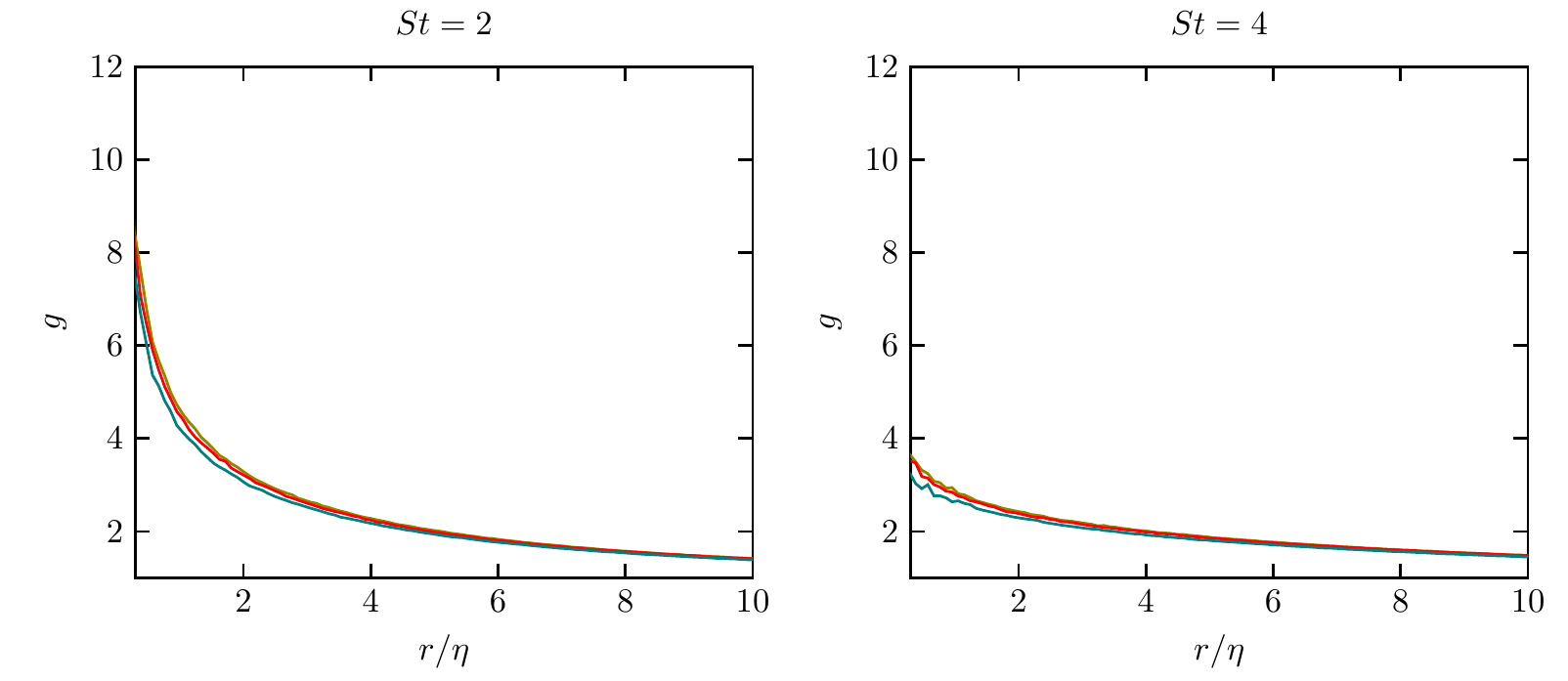}
    \includegraphics[scale=0.75]{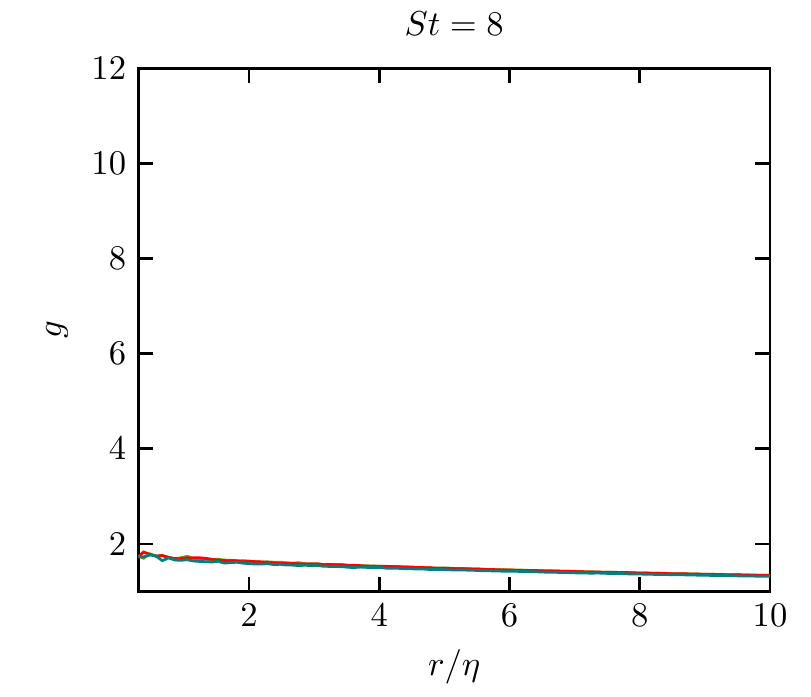}
    \caption{Radial distribution function of the one-way coupling simulations for different Stokes numbers and different thickness constants $\alpha$ of the interpolation of the subgrid-scale velocity between the sub-domains in forced HIT with the flow parameters given in table \ref{tab:ParametersSinglePhase}.}
    \label{fig:rdfalpha}
\end{figure}

Figure \ref{fig:rdfalpha} shows the radial distribution function of the one-way coupled simulations for different Stokes numbers and varying parameter $\alpha$ in the interpolation of the modeled subgrid-scale velocity between the sub-domains. The parameter $\alpha$ is reduced and increased by a factor of two relative to the value $\alpha=40$ that is used in the present study, respectively. It can be observed that even for this relatively wide parameter range the radial distribution functions almost coincide. The particle clustering is thus essentially independent of $\alpha$ for the considered range of values.
\end{document}